\documentclass[twocolumn]{aastex631}

\usepackage{booktabs}
\usepackage{multirow}
\let\tablenum\relax
\usepackage{siunitx}
\usepackage{amsmath}

\begin{document}

\title{A Closer Look at the Dynamical State of the High-redshift Galaxy Cluster SPT-CL J2215-3537}

\author[0000-0003-0307-7428]{Vinicius S. Bessa}
\affiliation{Observatório Nacional, 
Rua General José Cristino, 77, Bairro Imperial de São Cristovão 
Rio de Janeiro, RJ 20921400, Brazil}

\author[0000-0003-1477-3453]{Renato A. Dupke}
\affiliation{Observatório Nacional, 
Rua General José Cristino, 77, Bairro Imperial de São Cristovão 
Rio de Janeiro, RJ 20921400, Brazil}

\affiliation{Department of Astronomy, University of Michigan, 311 West Hall, 1085 South University Ave., Ann Arbor, MI 48109-1107}

\author[0000-0002-6090-2853]{Yolanda Jim\'enez-Teja}
\affiliation{Instituto de Astrof\'isica de Andaluc\'ia--CSIC, Glorieta de la Astronom\'ia s/n, E--18008 Granada, Spain}

\affiliation{Observatório Nacional, 
Rua General José Cristino, 77, Bairro Imperial de São Cristovão 
Rio de Janeiro, RJ 20921400, Brazil}

%\collaboration{20}{(AAS Journals Data Editors)}

%% Note that the \and command from previous versions of AASTeX is now
%% depreciated in this version as it is no longer necessary. AASTeX 
%% automatically takes care of all commas and "and"s between authors names.

%% AASTeX 6.31 has the new \collaboration and \nocollaboration commands to
%% provide the collaboration status of a group of authors. These commands 
%% can be used either before or after the list of corresponding authors. The
%% argument for \collaboration is the collaboration identifier. Authors are
%% encouraged to surround collaboration identifiers with ()s. The 
%% \nocollaboration command takes no argument and exists to indicate that
%% the nearby authors are not part of surrounding collaborations.

%% Mark off the abstract in the ``abstract'' environment. 
\begin{abstract}

We present a comprehensive reanalysis of the dynamical state of the high-redshift galaxy cluster SPT-CL J2215-3537 (z = 1.16), using the full set of available Chandra observations to characterize the thermodynamic and morphological properties of the intracluster medium. Although previously identified as one of the most distant dynamically relaxed systems based on X-ray morphological statistics, we find compelling evidence that SPT-CL J2215-3537 displays some level of dynamical activity. 
This includes temperature anisotropies consistent with
the first detection of a pair of core-sloshing cold fronts at $z>1$. We identify a ghost cavity candidate and estimate its mechanical power as $\log_{10}(P_{\text{cav}}/10^{42} \text{erg s}^{-1})=2.66\pm0.23$, confirming that radiative cooling strongly exceeds active galactic nucleus feedback heating. We show that SPT-CL J2215-3537 is likely in a short transient phase preceding the onset of a self-regulated cooling-feedback cycle. We recalculate traditional X-ray morphological parameters and discuss how non-self-similar evolution of parameters sensitive to the surface brightness cuspiness can bias dynamical classifications at high redshift.

\end{abstract}

%% Keywords should appear after the \end{abstract} command. 
%% The AAS Journals now uses Unified Astronomy Thesaurus concepts:
%% https://astrothesaurus.org
%% You will be asked to selected these concepts during the submission process
%% but this old "keyword" functionality is maintained in case authors want
%% to include these concepts in their preprints.

%% From the front matter, we move on to the body of the paper.
%% Sections are demarcated by \section and \subsection, respectively.
%% Observe the use of the LaTeX \label
%% command after the \subsection to give a symbolic KEY to the
%% subsection for cross-referencing in a \ref command.
%% You can use LaTeX's \ref and \label commands to keep track of
%% cross-references to sections, equations, tables, and figures.
%% That way, if you change the order of any elements, LaTeX will
%% automatically renumber them.
%%
%% We recommend that authors also use the natbib \citep
%% and \citet commands to identify citations.  The citations are
%% tied to the reference list via symbolic KEYs. The KEY corresponds
%% to the KEY in the \bibitem in the reference list below. 

\section{Introduction} \label{sec:intro}

As the most recently formed gravitationally bound objects in the Universe, galaxy clusters represent a unique laboratory for probing the nonlinear regime of large-scale structure formation, providing precise constraints on the underlying cosmological model. For instance, the abundance of clusters as a function of their mass, $dN/dM$, is sensitive to the matter density, $\Omega_m$, and to the amplitude of matter fluctuations on a scale of $8 \, h^{-1} \, \mathrm{Mpc}$, $\sigma_8$ \citep{1974ApJ...187..425P, 1993ApJ...407L..49B}. The distribution of clusters in redshift space, $dN/dz$, carries information on the dark energy equation-of-state parameter, $w_0$, through its dependence on the growth factor, $D_z$ and the cosmological volume element $dV/dz$ \citep{2001ApJ...553..545H}. Furthermore, due to the dominance of gravitational effects, it is expected that the gas mass fraction, $f_{\mathrm{gas}} = M_{\mathrm{gas}} / M_{\mathrm{tot}}$, in massive virialized clusters  $\gtrsim r_{2500}$ resembles the cosmic baryon fraction, $\Omega_b / \Omega_m$ \citep{1993Natur.366..429W, 1996PASJ...48L.119S,2002MNRAS.334L..11A}, and that, in this regime, it is fairly independent of redshift and halo mass \citep{2023A&A...675A.188A,2024arXiv241116555P}. The  $f_{\mathrm{gas}}$ method has been used, in combination with external priors, to place constraints on  \( \Omega_m \), \( \Omega_\Lambda \), and \( w \) \citep{2002MNRAS.334L..11A, 2009A&A...501...61E,2014MNRAS.440.2077M,2022MNRAS.510..131M}, providing independent evidence for cosmic acceleration. \\

The primary limitation of cluster counts and the  $f_{\mathrm{gas}}$ method as cosmological probes is the accurate determination of cluster masses. Generally, estimating the total halo mass requires the use of observable proxies, such as the X-ray emission from the intracluster medium (ICM), the weak lensing (WL) shear, or the Sunyaev-Zel'dovich (SZ) effect signal. The X-ray-derived mass typically relies on the assumptions of spherical symmetry and hydrostatic equilibrium (HSE) between the thermal gas and the gravitational potential, introducing a bias due to non-thermal pressure support from bulk motions, active galactic nuclei (AGN) feedback and turbulence, which can underestimate the total mass by tens of percent depending on the dynamical state of the gravitating material \citep{2007ApJ...655...98N,2009ApJ...705.1129L}. Many studies indicate that the hydrostatic mass bias is significantly reduced for derived masses at $r_{2500}$ \citep{1996ApJ...469..494E,2006MNRAS.369.2013R,2008MNRAS.384.1567M}. Moreover, several analytical corrections and calibrations have been proposed to systematically account for non-thermal components and recover true halo masses (e.g., \citealt{2004MNRAS.351..237R,2012ApJ...751..121N,2014MNRAS.443.1973V,2016MNRAS.457.1522A}). However, this scenario becomes significantly more complex for dynamically active systems (e.g., those experiencing recent or ongoing merger events). \cite{2014ApJ...792...25N} demonstrated through the analysis of 62 clusters from a high-resolution cosmological simulation that the hydrostatic bias in current/post-merger clusters can vary strongly with radius, becoming either highly positive or negative depending on the position of the infalling structure and making general mass correction schemes impractical. Additionally, an irreducible bias arises from gas acceleration components, which becomes non-negligible in unrelaxed systems, particularly in the high-redshift regime where mergers are more frequent. Even though the WL method is generally considered to be independent of dynamical assumptions, since its signal is proportional to the projected potential of the system, breaking the mass-sheet degeneracy still requires the adoption of a density analytical profile, which can be unsuitable for merging clusters \citep{2023ApJ...945...71L}. Additionally, the SZ-derived masses also inherit dependence on the cluster's dynamical state through their calibration with X-ray observables. One can avoid the strong dependence on the dynamical state by using the caustic technique to estimate the cluster's mass profile from kinematic information out to beyond $R_{200}$ \citep{1997ApJ...481..633D,1999MNRAS.309..610D,2016MNRAS.461.4182M}. However, in order to estimate the escape velocity profile, one needs a sufficiently high number of member galaxy redshifts, with at least $\sim100$ well constrained redshifts in the outskirts \citep{2009arXiv0901.0868D}, which limits the applicability of the method for fainter targets.  \\

Therefore, precise dynamical state classification is critical for cosmological applications, as mostly relaxed clusters provide sufficiently unbiased halo mass estimates. In the absence of high-quality spectroscopic data, the dynamical state of clusters may be inferred from the morphological regularity of the ICM X-ray emission. The pioneering work of \cite{2002MNRAS.334L..11A}, which established some of the first cosmological constraints from  $f_{\mathrm{gas}}$ measurements, selected relatively relaxed systems based on the sharpness of X-ray surface brightness peaks, isophote shapes, and isophote centroid variations, in addition to optical morphologies and gravitational lensing data, when available. Further studies \citep{2004MNRAS.353..457A,2008MNRAS.383..879A} extended the samples to X-ray luminous clusters in the redshift range $\sim 0.05 < z < \sim 1$, refining the selection criteria to include central cooling times and several other proxies for high-redshift systems with intermediate features. Many other works employed X-ray morphological parameters to trace substructures/inhomogenities in galaxy clusters \citep{2005ApJ...624..606J,2007A&A...467..485H,2015A&A...575A.127P,2020MNRAS.495..705Z,10.1093/mnras/staa2363}. This is a natural step, since the morphology parameters are directly calculated from the X-ray imaging, without requiring spectral modeling to derive the physical properties of the ICM. This allows for an automated approach to classify the increasing number of intermediate and high-redshift clusters currently detected via SZ and X-ray techniques prior to the upcoming powerful spectro-imagers such as AXIS (\href{https://axis.umd.edu/}{https://axis.umd.edu/}) \citep{2024Univ...10..273R} and NewAthena (\href{https://www.cosmos.esa.int/web/athena}{https://www.cosmos.esa.int/web/athena}) in the next decade, and also by the current large scale surveys such as J-PAS (\href{www.j-pas.org}{www.j-pas.org/}) and LSST (\href{www.lsst.org/}{www.lsst.org/}). \\

Building upon this scheme, \cite{2015MNRAS.449..199M} developed the Symmetry–Peakiness–Alignment (SPA) criteria, a set of X-ray morphological parameters, designed to be applicable to large datasets spanning a wide range of redshifts, masses, and data qualities. With this framework, \cite{2022MNRAS.510..131M} included the high-redshift ($z = 1.16$) cluster SPT-CL J2215-3537 (SPT2215 hereafter) \citep{2020ApJS..247...25B} in their extended sample of massive relaxed clusters suitable for $f_{\mathrm{gas}}$ cosmology. \cite{2023ApJ...947...44C} identified it as the most distant relaxed cool core cluster at the time, providing evidence for a highly efficient central cooling and a remarkably high star formation rate in the brightest cluster galaxy (BCG). However, recent studies have been reevaluating, using independent methods, the dynamical state of galaxy clusters previously classified as relaxed based on morphological parameters. For instance, \cite{2023A&A...676A..39J} performed a multiwavelength study of the high-redshift ($z=0.97)$ cluster SPT-CLJ0615-5746, which was also one of the objects included in the extended sample of massive relaxed clusters, aiming to probe the dynamical state of the system through the intracluster light  fraction ($f_{\text{ICL}}$) and the X-ray emission of the ICM. The work found a distribution of rest-frame $f_{\text{ICL}}$ inconsistent with a relaxed system. Aditionally, the temperature map derived from the X-ray analysis showed signatures of multiple mergers. \\

%Following this trend, \cite{hyeonghan2025direct} performed a weak-lensing analysis in the Perseus cluster ($z=0.0179$) and detected a nearby subcluster halo linked by a mass bridge, confirming the scenario of a past major merger. \\ NOT SURE IF THIS FITS THE CONTEXT  

In this work, we re-examine the dynamical state of SPT2215 with newly released Chandra data and discuss the effectiveness of morphological parameters as cluster relaxation proxies, particularly at high redshift. The paper is organized as follows. In Section \ref{sec:spec}, we describe the X-ray data reduction and perform a spectral analysis to derive thermodynamic profiles and maps. In Section \ref{sec:imag}, we analyze the merged X-ray image to characterize the spatial properties of the ICM gas, additionally computing commonly used morphological parameters. Section \ref{sec:discu} discusses the results of our combined analysis and proposes consistent scenarios for SPT2215's dynamical history. Finally, Section \ref{sec:5} summarizes our findings and present future research directions based on our conclusions.

\section{X-Ray Spectral Analysis} \label{sec:spec}

All relevant quantities were calculated considering a standard flat $\Lambda$ cold dark matter cosmology, with $H_0 = 70 \,\text{km} \,\text{s}^{-1} \text{Mpc}^{-1}, \quad \Omega_m = 0.3, \quad \text{and} \quad \Omega_\Lambda = 0.7$. Unless otherwise noted, all quoted uncertainties represent 1$\sigma$ confidence intervals and, when needed, are propagated using the partial derivatives method considering the covariance matrices given by the fits. The science products were obtained in the usual way using the X-ray Data Analysis Software (CIAO 4.17, \citealt{2006SPIE.6270E..1VF}). We used all ten archived Chandra observations available: three were taken with ACIS-I (OBS-ID: 22653, 24614, 24615; PI: McDonald), and seven with ACIS-S  (OBS-ID: 25468, 25902, 25903, 25904, 26244, 26372, 27614; PI: Mantz). The individual observations were downloaded and reprocessed using up-to-date calibration data from CALDB version 4.11.6 with additional background cleaning for the VFAINT telemetry mode. The reprocessed events files were filtered to the 0.5-7 keV band and the ACIS-I3 or ACIS-S3 chips were selected according to the observation. The exposure maps were created with the \textit{fluximage} script using source characteristic spectral weights. Point sources were removed using the \textit{wavedetect} tool followed by an additional eye check. Bad time intervals were identified through the source
lightcurves in order to eliminate flares and create cleaned event files. The total filtered exposure time was 188.12 ks.  \\

\subsection{Radial Thermodynamic Profiles} \label{sec:radial}

Annuli regions around the BCG position ($\alpha = 22^{\text{h}}15^{\text{m}}03.\!^{\text{s}}9306, \quad \delta = -35^\circ37'17.''885$) were selected. The correspondent Blank-Sky files were used to account for the background contribution. The spectra of all reprojected observations were extracted using \textit{specextract} and then simultaneously fitted with a \textit{phabs*apec} model using Xspec version 12.14.1 \citep{1996ASPC..101...17A}. The abundances were fixed at $0.3 \, Z_{\odot}$, where $Z_{\odot}$ is the solar photospheric elemental abundance of \cite{1989GeCoA..53..197A}. Temperatures were tied to each other and the remaining parameters were kept independent. The hydrogen column was fixed at $nH = 1.01 \cdot 10^{20}$ cm$^{-2}$ according to the HI4PI survey \citep{2016A&A...594A.116H}. The deprojection of thermodynamic quantities was performed using the onion-peeling method under the assumption of spherical symmetry in concentric shells, similarly to the recipe presented in \cite{2004ApJ...609..638G}. The electron number density $n_e$ was derived from the deprojected normalization parameter $N$ of the spectral model,

 \begin{equation}
    n_e = 10^7 D_A(1+z) \sqrt{\frac{4\pi N}{0.83 V}},
\label{eq:ne}
 \end{equation}

where $D_A = 1701.686$ Mpc is the angular diameter distance, $z = 1.16$ is the redshift, $V$ is the considered shell volume, and $0.83$ term comes from the factor between the proton density and the electron density ($n_p=0.83n_e$). We additionally fit the surface brightness with a one-dimensional beta model, assuming the cooling function $\Lambda(k_BT,Z)$ parameterized by \cite{2001ApJ...546...63T}, allowing us to obtain an analytical curve for the electron number density, given by

 \begin{equation}
    n_e(r) = n_0\left[1+\left(\frac{r}{r_c}\right)^2\right]^{-3\beta/2}.
\label{eq:nbet}
 \end{equation}

 The core radius $r_c$, the beta parameter $\beta$ and the central surface brightness $S_0$ come directly from the fit. The central density is calculated using the following.

 \begin{equation}
    n_0 = \left[\frac{S_0}{\sqrt{\pi} r_c  \Lambda(k_BT,Z)} \frac{\Gamma(3 \beta)}{\Gamma (3\beta - 0.5)}\right]^{1/2},
\label{eq:entr}
 \end{equation}

where $\Gamma(x)$ represents the gamma function. With the temperature $T$ and the electron number density  $n_e$ profiles, we can easily compute the pseudopressure and pseudoentropy, which are, respectively, given by the following relations

 \begin{equation}
    P(r) = 1.92n_ekT,
\label{eq:press}
 \end{equation}

 \begin{equation}
    K(r) = kTn_e^{-2/3},
\label{eq:entr1}
 \end{equation}

 where $k$ is the Boltzmann constant. Assuming a compressionless thermal cooling, we can calculate the cooling time as

  \begin{equation}
    t_c(r) = \frac{3P}{2{n_e}^2\Lambda (k_BT,Z)}.
\label{eq:cool3}
 \end{equation}

\begin{figure}[ht!]
\plotone{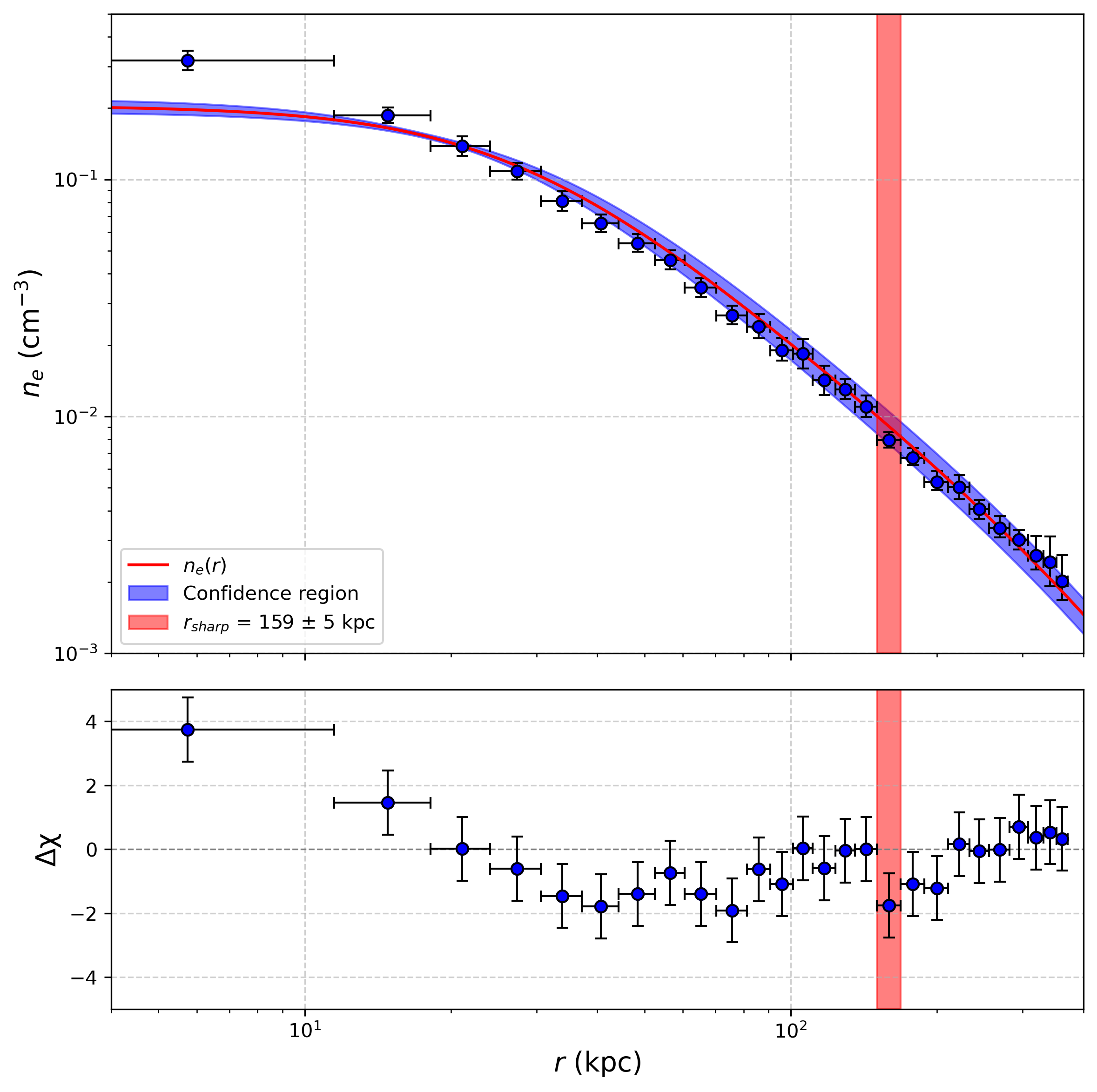}
\caption{
Top: Deprojected electron density profile. The blue dots were obtained from Equation~\ref{eq:ne}, while the red curve corresponds to the $\beta$-model applied to the surface brightness profile, whose optimal parameters are inserted in Equation~\ref{eq:nbet}.
Bottom: residual between the data and the model. The sharp decline observed in both the density profile and the residual around $r_{\mathrm{sharp}} \approx 160$ kpc / $19.4$ arcsec is associated to two symmetric cold fronts.
\label{fig:nepro}
}
\end{figure}

\begin{figure*}[ht!]
    \centering
    \includegraphics[width=\textwidth]{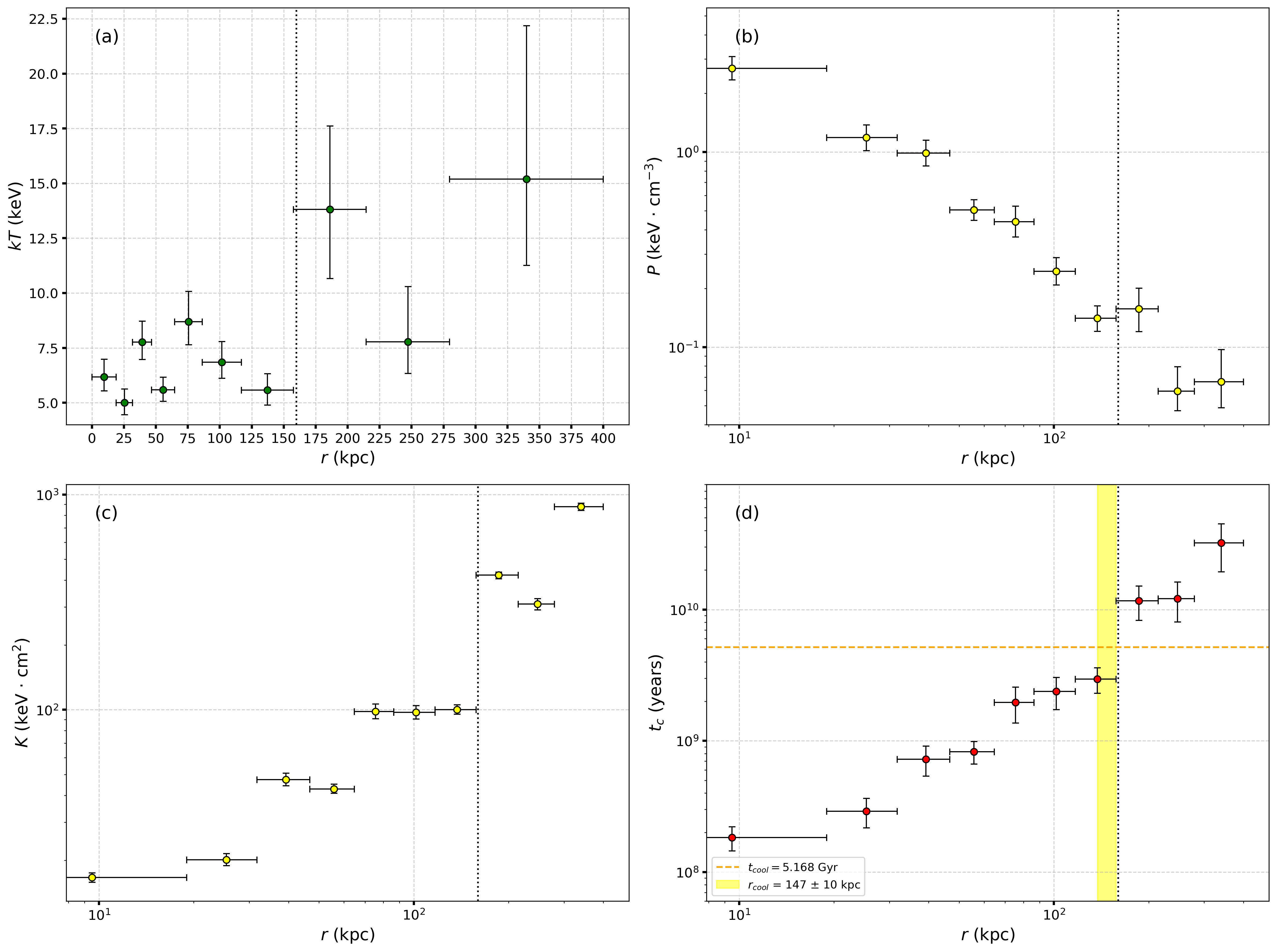}
    \caption{Radial profiles of various 3D thermodynamic quantities: (a) Temperature ($kT$), (b) Pressure ($P$), (c) Entropy ($K$), and (d) Cooling Time ($t_c$). The shaded region in (d) indicates the cooling radius $r_{cool}$. The dotted vertical line in all panels mark $r = 160$ kpc to indicate the approximate projected radius of the observed discontinuity.}
    \label{fig:radpro}
\end{figure*}

Figure~\ref{fig:nepro} shows the deprojected electron density profile both from the spectral modeling and from the $\beta$-model fit. The electron density exhibits a pronounced decline at $r_{\text{sharp}}=159\pm5$ kpc / $\sim19.3\pm0.6$ arcsec. In Figure \ref{fig:radpro} we show the correspondent radial profiles of the main 3D thermodynamic quantities. The feature in $r_{\text{sharp}}$ is further corroborated in the thermodynamic profiles, where we observe a steep rise in both temperature and entropy beyond the discontinuity, while the pressure remains nearly constant. All projected radial profiles consistently point toward a common feature: a contact discontinuity located at a projected radius of approximately $160$ kpc. The presented picture is consistent with the physical conditions expected for a cold front. Interestingly, the estimated cooling radius, $r_{\rm cool} = 147 \pm 10$ kpc / $17.8\pm1.2$ arcsec, defined here as the radius where the local cooling time matches the age of the Universe at the cluster's redshift ($\sim5.17$ Gyr), is very close to the location of the discontinuity. Another feature worth mentioning is the flat central temperature profile for regions below $r_{\rm cool}$. Although SPT2215 manifest clear properties of a well developed cool core (e.g., centrally peaked surface brightness, central entropy below $30$ keV$\cdot$cm$^2$, central cooling time below $1$ Gyr), its temperature profile deviates from the classical cool core behavior, lacking the expected continuous decline toward the cluster center.

As a consistency check, we verify that our temperature and emission measure profiles are equivalent to those presented by \cite{2023ApJ...947...44C} when using the same first three observations. By using all ten observations, we obtain a mean reduction in the uncertainties by a factor of $\sim2$ for both the temperature and normalization fits. This improvement allows us to choose finer radial bins (necessary to detect the discontinuity) and construct thermodynamic maps, which would not be achievable with the required statistical significance solely with the initial set of observations. The radial extent of our analysis ($\sim 430$ kpc / $48.5$ arcsec) was determined by identifying out to which radius the spectral fits no longer converge. We note that the source surface brightness becomes comparable to the background level at $\sim 580$ kpc /$70$ arcsec. Finally, we verify that the results we obtain using the rescaled blanksky files are consistent to those derived using a local background region.

\subsection{Projected Thermodynamic Maps}

In order to easily identify directions of interest and probe potential anisotropies in the physical properties of the ICM, we constructed maps of the projected thermodynamic parameters. To do so, we first applied a Weighted Voronoi Tesselations (WVT) algorithm to partition the field of view, ensuring a nearly constant S/N. The basic Voronoi Tesselations algorithm for spatial adpative binning is described in details in \cite{2003MNRAS.342..345C}.  Several applications of the WVT method to X-ray data are presented in \cite{2006MNRAS.368..497D}, as well as its advantages over other well-used algorithms such as the Quadtree \citep{2001MNRAS.325..178S}. Then, the spectrum of each region/bin is extracted and fitted with Xspec using the \textit{phabs*apec} model as before. The WVT algorithm naturally minimizes the sharpness of bin edges, but they are still discontinuous. To improve the visualization, we applied Universal Kriging interpolation using the PyKrige\footnote{The package documentation is available on \url{http://pykrige.readthedocs.io/}. The temperature and entropy maps used a spherical variogram model with linear drift terms, while the pressure map used a hole-effect variogram model, as the standard ones were unable to adequately capture the spatial structure of the data.} Python package (version 1.7.2; \citealt{pykrige}), which allows spatial smoothing while accounting for the underlying trends in the data.

\begin{figure*}[ht!]
    \centering
    \includegraphics[width=\textwidth]{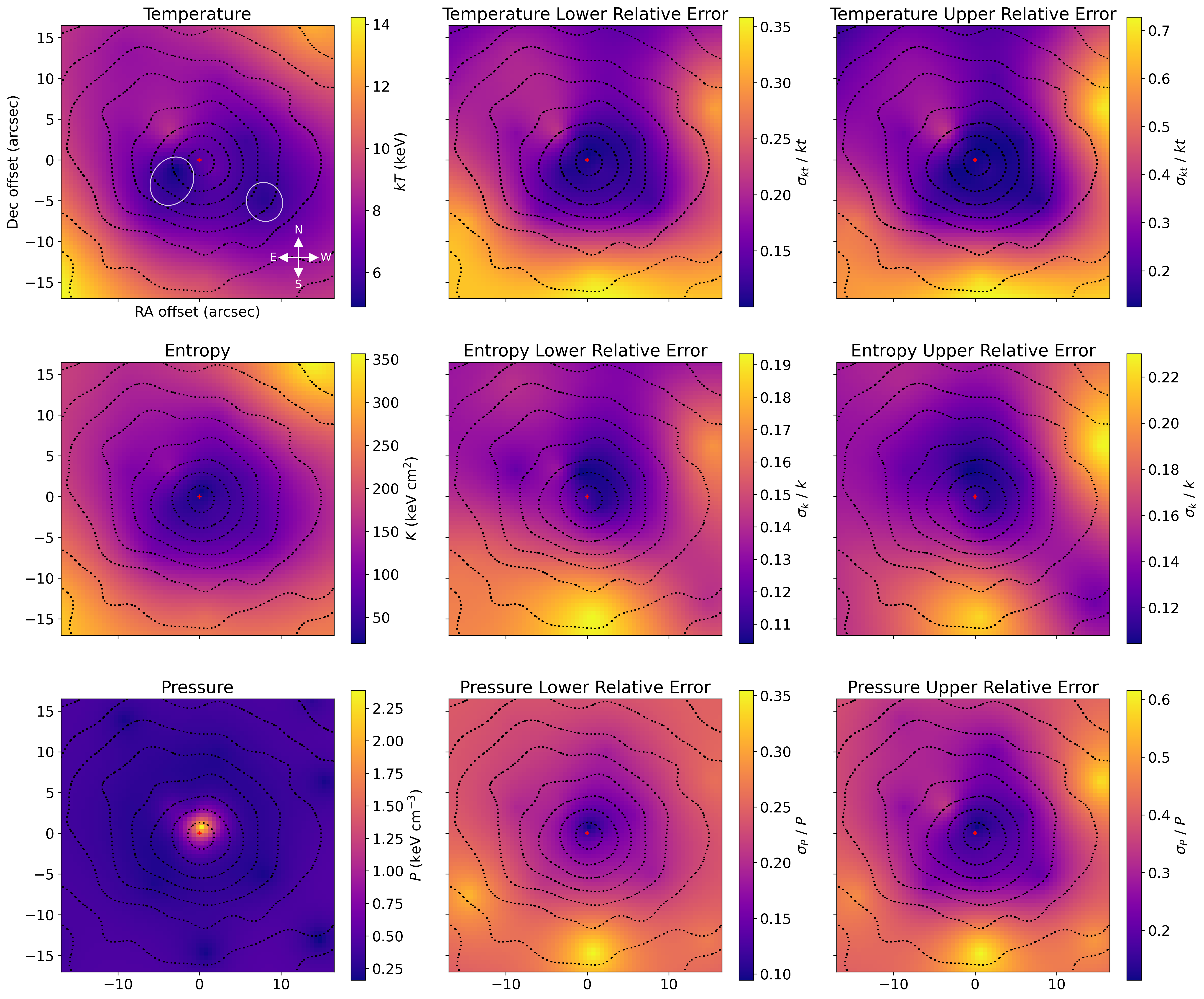}
    \caption{Projected thermodynamic maps (temperature, pseudo-entropy, and pseudo-pressure) derived using WVT adaptive binning followed by kriging interpolation, along with the corresponding lower and upper residual error maps. White ellipses in the temperature map indicate the approximate locations of colder clumps. The red cross marks the position of the BCG. Coordinates are shown as offsets from the BCG in arcseconds ($1 \text{ arcsec}= 8.25 \text{ kpc}$).}
    \label{fig:thermo_maps}
\end{figure*}

In Figure \ref{fig:thermo_maps} we present the maps of the projected thermodynamic quantities. It is important to note that because of the low S/N regime, these maps should be read qualitatively. The temperature map shows some interesting features. First, it appears that the coldest region is displaced from the BCG center by roughly $5$ arcsec ($\sim50$ kpc) to the southeast. We fit the spectra of a joint set of these peripheral cold clumps, indicated in Figure \ref{fig:thermo_maps} as white ellipses, and find a statistical significance of $1.8 \sigma$ that these regions are indeed colder than the center. The second feature is the apparent sudden temperature increase in the southeast and northwest directions approximately around $r_{cool}$, confirming the orientation of the discontinuities.  On the other hand, the entropy map exhibits a smoother distribution in the core, with its lowest value aligned with the cluster's center.  The pressure map suggests the presence of possible central substructures, with a higher pressure tail towards the NE direction.

\begin{deluxetable}{lccccc}
\tabletypesize{\scriptsize}
\tablewidth{0pt}
\tablecaption{Global Cluster Properties from Coarse Binning \label{tab:coarse}}
\tablehead{
\colhead{$r \pm \Delta r$} & \colhead{$kT$} & \colhead{$P$} & \colhead{$K$} & \colhead{$t_c$} & \colhead{$\chi^2_{\rm red}$} \\
\colhead{(kpc)} & \colhead{(keV)} & \colhead{(keV cm$^{-3}$)} & \colhead{(keV cm$^2$)} & \colhead{(10$^8$ yr)} & \colhead{}
}
\startdata
$9.5 \pm 9.5$   & $6.18^{+0.80}_{-0.64}$  & $2.69^{+0.40}_{-0.34}$ & $16.61^{+0.84}_{-0.82}$  & $1.83 \pm 0.38$   & 0.817 \\
$25.4 \pm 6.4$  & $5.00^{+0.62}_{-0.54}$  & $1.19^{+0.40}_{-0.17}$ & $20.11^{+1.36}_{-1.27}$  & $2.91 \pm 0.74$   & 0.699 \\
$39.2 \pm 7.5$  & $7.77^{+0.96}_{-0.80}$  & $0.99^{+0.16}_{-0.14}$ & $47.37^{+3.34}_{-3.06}$  & $7.25 \pm 1.87$   & 0.799 \\
$55.7 \pm 9.0$  & $5.59^{+0.57}_{-0.53}$  & $0.51^{+0.06}_{-0.06}$ & $42.89^{+2.18}_{-2.01}$  & $8.25 \pm 1.61$   & 0.721 \\
$75.5 \pm 10.4$ & $8.69^{+1.38}_{-1.05}$  & $0.44^{+0.09}_{-0.07}$ & $98.16^{+8.08}_{-7.42}$  & $19.62 \pm 5.99$  & 0.761 \\
$101.6 \pm 13.1$& $6.85^{+0.94}_{-0.73}$  & $0.24^{+0.04}_{-0.04}$ & $97.42^{+7.10}_{-6.80}$  & $23.81 \pm 6.52$  & 0.777 \\
$137.2 \pm 18.7$& $5.57^{+0.75}_{-0.68}$  & $0.14^{+0.02}_{-0.02}$ & $100.07^{+5.30}_{-4.84}$ & $29.50 \pm 6.44$  & 0.781 \\
$186.0 \pm 27.4$& $13.82^{+3.79}_{-3.16}$ & $0.16^{+0.04}_{-0.04}$ & $422.34^{+14.30}_{-15.91}$& $116.77 \pm 33.93$& 0.872 \\
$247.1 \pm 33.5$& $7.77^{+2.52}_{-1.44}$  & $0.06^{+0.02}_{-0.01}$ & $309.64^{+18.38}_{-18.74}$& $121.07 \pm 40.73$& 0.925 \\
$339.8 \pm 58.1$& $15.20^{+6.98}_{-3.94}$ & $0.07^{+0.03}_{-0.02}$ & $878.24^{+32.58}_{-33.80}$& $321.72 \pm 128.23$& 0.943 \\
\enddata
\tablecomments{Radial profiles of global thermodynamic properties derived from spectral modeling. 
(1) Radius; 
(2) Temperature; 
(3) Pseudo-pressure; 
(4) Pseudo-entropy;
(5) Cooling time
(6) Reduced $\chi^2$ of spectral fit.}
\end{deluxetable}

\subsection{Directional Thermodynamic Profiles} \label{sec:direc}

\begin{figure}[h]
\plotone{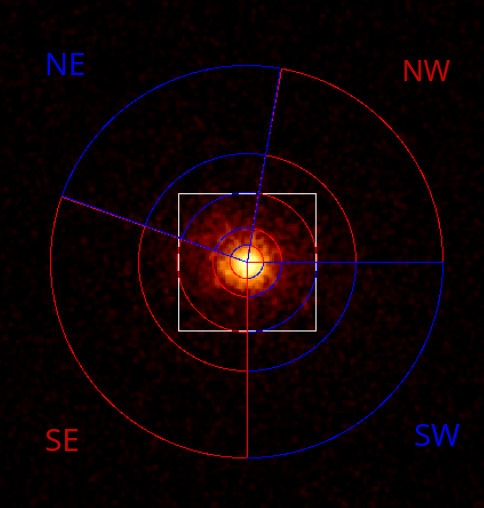}
\caption{Directional sectors and central region used in the analysis. The $17'' \times 17''$ white box marks the area of the thermodynamic maps. Sectors: NE ($\ang{80}$–$\ang{160}$), SE ($\ang{160}$–$\ang{270}$), NW ($\ang{0}$–$\ang{80}$), SW ($\ang{270}$–$\ang{360}$).
\label{fig:ds9dirl}}
\end{figure}

\begin{figure*}[ht!]
    \centering
    \includegraphics[width=\textwidth]{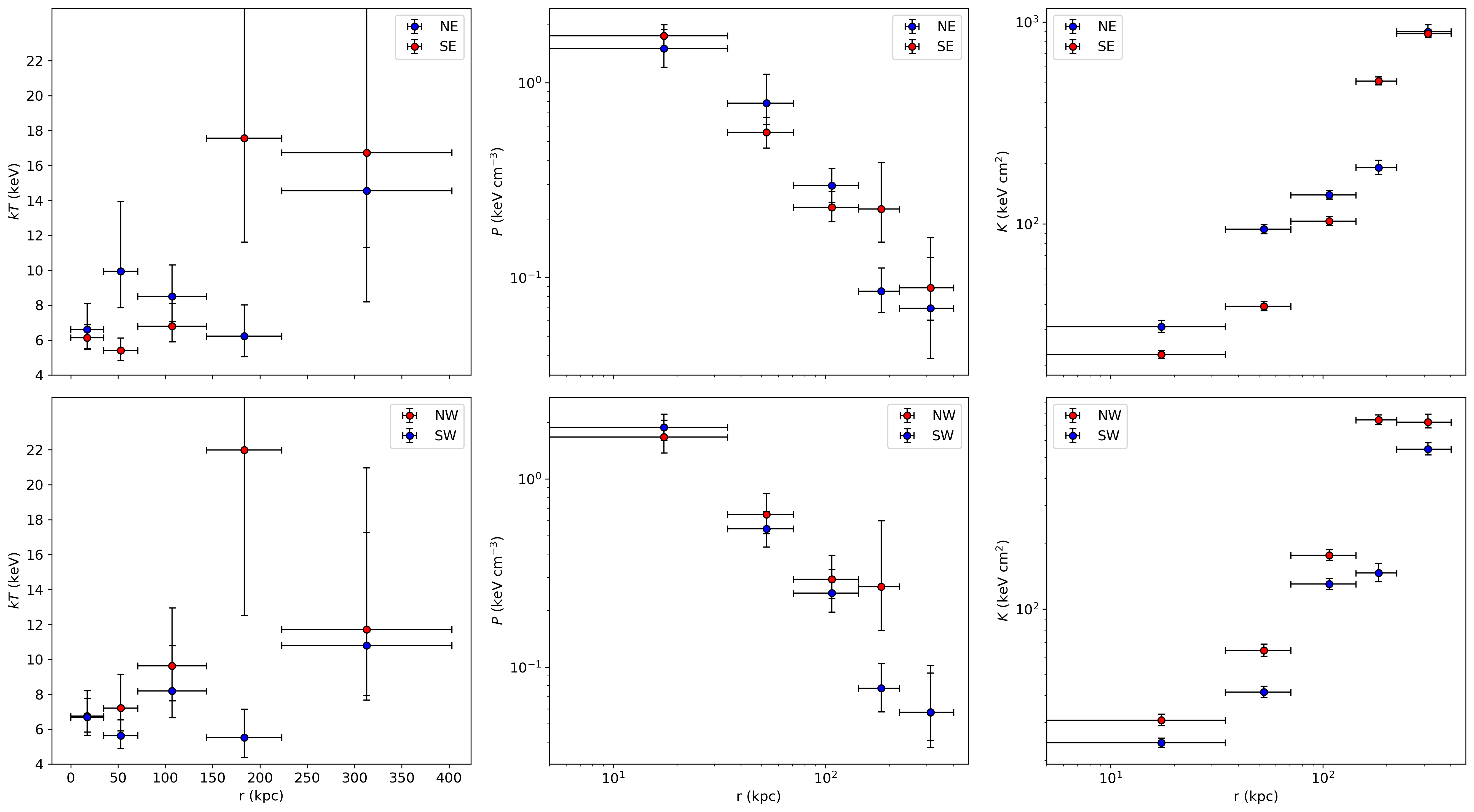}
    \caption{Directional projected thermodynamic profiles extracted from sector-shaped annuli. The upper panels compare the NE and SE directions, while the lower panels show the NW versus SW directions.}
    \label{fig:dirpro}
\end{figure*}

Based on the spatial features revealed in the thermodynamic maps, we defined concentric annular regions along four specific directions: northwest (NW), northeast (NE), southeast (SE), and southwest (SW), as indicated in Figure~\ref{fig:ds9dirl}. We then applied the same procedure described in Section~\ref{sec:radial} to derive directional thermodynamic profiles. The results, presented in Figure~\ref{fig:dirpro}, are consistent with the trends observed in the thermodynamic maps, revealing anisotropies in the central regions. While the SE-NW axis shows a relatively smooth temperature decline to the center, the temperature profile in the SW-NE axis has clear temperature dips and entropy flattening at $r\approx 180$ kpc. The geometry of the opposed dipolar discontinuities seen in the temperature map of Figure \ref{fig:thermo_maps} is consistent with cold fronts viewed close to the plane of the sky. This makes it unlikely that the observed flattening of these profiles are due to projection effects. We discuss the possible origin of this feature in Section \ref{sec:4.3}.  Notably, at the radial bin corresponding to the
identified discontinuity at $r_{\text{sharp}} \approx 160$ kpc, the profiles along the SE–NW axis exhibit similar behavior, characterized by coincident temperature and entropy enhancements as well as a flat pressure profile. These features confirm the presence of two aligned cold fronts along this axis.

\begin{deluxetable}{lccccc}
\tabletypesize{\scriptsize}
\tablewidth{0pt}
\tablecaption{Directional Cluster Properties \label{tab:fine}}
\tablehead{
\colhead{Dir.} & 
\colhead{$r \pm \Delta r$} & 
\colhead{$kT$} & 
\colhead{$P$} & 
\colhead{$K$} & 
\colhead{$\chi^2_{\rm red}$} \\
\colhead{} & 
\colhead{(kpc)} & 
\colhead{(keV)} & 
\colhead{(keV cm$^{-3}$)} & 
\colhead{(keV cm$^2$)} & 
\colhead{} 
}
\startdata
% NE Direction
NE & $17.4 \pm 17.4$ & $6.61^{+1.49}_{-1.16}$ & $1.50^{+0.38}_{-0.30}$ & $31.00^{+2.31}_{-1.93}$ & 0.944 \\
NE & $52.8 \pm 18.1$ & $9.94^{+3.99}_{-2.08}$ & $0.78^{+0.32}_{-0.18}$ & $94.18^{+5.26}_{-4.82}$ & 0.746 \\
NE & $107.2 \pm 36.3$ & $8.50^{+1.80}_{-1.45}$ & $0.30^{+0.07}_{-0.05}$ & $139.09^{+7.24}_{-6.17}$ & 0.814 \\
NE & $183.4 \pm 43.6$ & $6.23^{+1.79}_{-1.18}$ & $0.09^{+0.03}_{-0.02}$ & $190.18^{+16.62}_{-14.80}$ & 0.724 \\
NE & $313.0 \pm 99.3$ & $14.55^{+11.83}_{-6.36}$ & $0.07^{+0.06}_{-0.03}$ & $895.58^{+70.76}_{-53.51}$ & 0.919 \\
\hline
% SE Direction
SE & $17.4 \pm 17.4$ & $6.14^{+0.75}_{-0.63}$ & $1.74^{+0.25}_{-0.22}$ & $22.57^{+2.31}_{-1.93}$ & 0.699 \\
SE & $52.8 \pm 18.1$ & $5.41^{+0.72}_{-0.58}$ & $0.56^{+0.11}_{-0.10}$ & $39.09^{+2.13}_{-1.97}$ & 0.747 \\
SE & $107.2 \pm 36.3$ & $6.80^{+1.30}_{-0.89}$ & $0.23^{+0.05}_{-0.04}$ & $103.38^{+5.45}_{-5.34}$ & 0.788 \\
SE & $183.4 \pm 43.6$ & $17.56^{+13.27}_{-5.94}$ & $0.23^{+0.04}_{-0.04}$ & $508.97^{+25.37}_{-21.77}$ & 0.767 \\
SE & $313.0 \pm 99.3$ & $16.73^{+14.16}_{-5.43}$ & $0.09^{+0.04}_{-0.02}$ & $875.74^{+44.61}_{-40.46}$ & 0.861 \\
\hline
% NW Direction
NW & $17.4 \pm 17.4$ & $6.75^{+1.45}_{-1.10}$ & $1.67^{+0.38}_{-0.30}$ & $30.72^{+2.06}_{-1.80}$ & 0.867 \\
NW & $52.8 \pm 18.1$ & $7.21^{+1.92}_{-1.31}$ & $0.65^{+0.11}_{-0.09}$ & $64.46^{+4.47}_{-3.92}$ & 0.700 \\
NW & $107.2 \pm 36.3$ & $9.62^{+3.32}_{-2.00}$ & $0.29^{+0.10}_{-0.06}$ & $176.90^{+10.53}_{-9.40}$ & 0.755 \\
NW & $183.4 \pm 43.6$ & $21.98^{+28.48}_{-9.47}$ & $0.27^{+0.33}_{-0.11}$ & $745.28^{+39.20}_{-37.50}$ & 0.693 \\
NW & $313.0 \pm 99.3$ & $11.72^{+9.24}_{-4.05}$ & $0.06^{+0.04}_{-0.02}$ & $727.54^{+64.57}_{-44.37}$ & 0.913 \\
\hline
% SW Direction
SW & $17.4 \pm 17.4$ & $6.69^{+1.08}_{-0.85}$ & $1.88^{+0.33}_{-0.27}$ & $24.12^{+1.24}_{-1.15}$ & 0.772 \\
SW & $52.8 \pm 18.1$ & $5.63^{+0.90}_{-0.74}$ & $0.54^{+0.13}_{-0.11}$ & $41.34^{+2.63}_{-2.32}$ & 0.801 \\
SW & $107.2 \pm 36.3$ & $8.18^{+2.60}_{-1.53}$ & $0.25^{+0.08}_{-0.05}$ & $130.28^{+8.02}_{-7.42}$ & 0.686 \\
SW & $183.4 \pm 43.6$ & $5.52^{+1.63}_{-1.13}$ & $0.08^{+0.03}_{-0.02}$ & $146.56^{+16.03}_{-12.99}$ & 0.851 \\
SW & $313.0 \pm 99.3$ & $10.80^{+6.47}_{-2.88}$ & $0.06^{+0.04}_{-0.02}$ & $546.30^{+38.18}_{-33.34}$ & 0.987 \\
\enddata
\tablecomments{Azimuthal variations in thermodynamic properties. 
(1) Direction sector; 
(2) Radius; 
(3) Temperature; 
(4) Pseudo-pressure; 
(5) Pseudo-entropy; 
(6) Reduced $\chi^2$ of spectral fit.}
\end{deluxetable}

\section{X-Ray Imaging Analysis} \label{sec:imag}

To visually inspect the ICM anisotropies hinted by the thermodynamic maps, we reproject the observations to a common tangent point and create a merged, exposure-corrected image. Again, we use the correspondent blanksky files to subtract the scaled backgrounds from the science images.  
\subsection{Residual map} \label{sec:resid}

We start by adaptatively smoothing the merged image using the CIAO tool \textit{dmimgadapt} with a gaussian kernel, demanding 16 counts per kernel and a minimum and maximum scale of, respectively, 0.5 and 15 pixels. Then, we proceed to model the radial distribution of the surface brightness with an elliptical 2D double-$\beta$ model, where each component is defined by 

\begin{equation}
    S(x_i,y_i) = S(r) = A\left[1+\left(\frac{r}{r_c}\right)^2\right]^{-3\beta+0.5} + C,
    \label{eq:2dbeta}
\end{equation}

where

\begin{equation}
    r(x_i,y_i) = \frac{\sqrt{x^2(1-e)^2+y^2}}{1-e},
\end{equation}

and

\begin{equation}
\begin{aligned}
x &= (x_i - x_0)\cos\theta + (y_i - y_0)\sin\theta, \\
y &= (y_i - y_0)\cos\theta - (x_i - x_0)\sin\theta.
\end{aligned}
\end{equation}

\begin{figure*}[ht!]
    \centering
    \includegraphics[width=\textwidth]{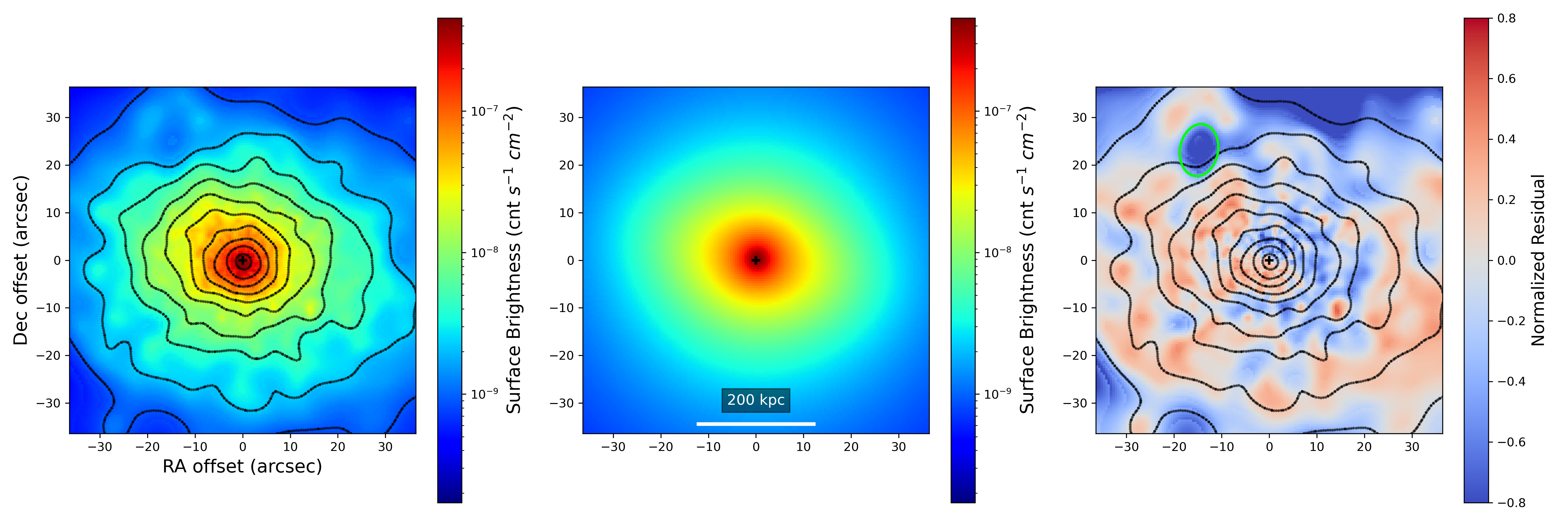}
    \caption{Left: Adaptively smoothed X-ray surface brightness map with overlaid contours. Middle: Best-fit double two-dimensional $\beta$-model (Equation \ref{eq:2dbeta}); the optimal parameters are listed in Table \ref{tab:2dbeta}. Right: Normalized residual map with the surface brightness contours. The green ellipse-centered at $\alpha = 22^{\text{h}}15^{\text{m}}05.^{\text{s}}1467$, $\delta = -35^\circ36'54.''829$, with semi-axes of $5.5\pm1$ and $4.0\pm1$ arcseconds and a position angle of $350^\circ$-encloses the X-ray cavity candidate. The black cross marks the position of the BCG. Coordinates are shown as offsets from the BCG in arcseconds ($1 \text{ arcsec}= 8.25 \text{ kpc}$).}
    \label{fig:residual}
\end{figure*}

Here, ($x_0$,$y_0$) is the centroid of the cluster, $e$ is the ellipticity and $\theta$ is the orientation of the major axis relative to the north in the anti-clockwise direction. The initial values of the cluster shape parameters were calculated using the method of moments of inertia, as described in \cite{1991MNRAS.249..662P}, using $S(r)$ as the generic function. To optimize the parameters we perform a Levenberg-Marquardt optimization to refine the near-optimal solutions and properly estimate the uncertanties. The best fit parameters for the surface brightness model are presented in Table \ref{tab:2dbeta}. Then, we calculate the normalized residual with $(\text{Data}-\text{Model})/\text{Data}$. The residual map in Figure \ref{fig:residual} exhibits a relatively low overall amplitude ($\pm0.25$), but reveals a pronounced dipolar pattern within 15 arcseconds of the BCG. The directional variation in enhanced emission across the cluster radii suggests that this anisotropy is unlikely to be caused by projection effects from a triaxial structure. Instead, it likely arises from a non-symmetric distribution of cold gas, possibly driven by core sloshing. Indeed, the directional surface brightness profiles normalized by the azimuthally averaged brightness shows the oscillatory behavior characteristic of spiral-shaped cold gas structures near the cluster core. These features have been detected in X-ray imaging \citep{2004ApJ...616..178C}, temperature maps \citep{2005MNRAS.360L..20F} and even in chemical enrichment distributions \citep{2007ApJ...671..181D}, being also naturally reproduced in simulations \citep{2006ApJ...650..102A}. \\

Another finding from the $\beta$-model subtraction is a statistically significant extended depression (residual $< -0.5$ and local brightness dimming $\sim 25\%$). If we consider that this is a jet-driven X-ray cavity, we can model the bubble as a prolate ellipsoid with semi-axes $r_1$, $r_2$, and $r_3 = r_2$, where $r_1$ and $r_2$ correspond to the semi-major and semi-minor axes of the projected ellipse, respectively. The bubble's physical volume is therefore given by $V = (4/3)\pi r_1 r_2^2$. Following standard approaches, we assume this volume contains a population of relativistic electrons in pressure equilibrium with the surrounding ICM. From the cavity's projected elliptical shape, we estimate a volume of ($2.07\pm1.10$)$\times 10^5$ kpc$^3$. From the spatially resolved thermodynamic information, we estimate that the necessary energy to inflate such bubble is  ($2.73\pm1.86$)$\times10^{60}$ erg. Considering that the projected distance between the BCG and the cavity's outermost edge is $\sim255$ kpc, we calculate a sound-crossing time of $188\pm22$ Myr. Our calculation of the buoyancy time for this same separation yields $t_{\text{buoy}}=236$ Myr, however, it is hard to say if this value is reliable, since estimating the buoyancy requires a well-constrained enclosed mass profile, in order for one to compute the local acceleration. In our case, we try to reproduce \cite{2023ApJ...947...44C} methodology by fitting the deprojected temperature profile with the \cite{2006ApJ...640..691V} model. Nevertheless, we find that the shape of the profile is highly sensitive to assumptions about the deprojected temperature structure (e.g., the inclusion of the discontinuity bin, assuming a stable profile beyond $\sim 430$ kpc). For this reason, we consider the sound-crossing time to be a more robust choice. In any case, from the enclosed-mass profile we obtain by solving the hydrostatic equation, we estimate relevant scale radii to be $R_{2500}\sim350$ kpc, $R_{500}\sim850$ kpc and $R_{180}\sim1480$ kpc. Considering the sound-crossing time as the characteristic timescale for the bubble propagation, our final estimative for the jet power responsible for this possible cavity is $\log_{10}(P_{\text{cav}}/10^{42} \text{erg s}^{-1})=2.66\pm0.23$, consistent with the value reported by \cite{2023ApJ...947...44C} using the $P_{\text{cav}}$-$L_{\text{cool}}$ scaling relation \citep{2010ApJ...720.1066C}. Using $t_{\text{buoy}}$ does not heavily affect the inferred cavity power, as a difference of $\sim25\%$ in the timescale results in a change of less than $0.1$ dex in the cavity power, well within the original uncertainty. In Section \ref{sec:4.2}, we further discuss the feasibility of such a bubble given its observed characteristics, as well as its implications for the cluster's dynamical history.

\begin{deluxetable}{lcc}
\tabletypesize{\scriptsize}
\tablewidth{0pt}
\tablecaption{Double Elliptical $\beta$-Model Fit Optimal Parameters \label{tab:2dbeta}}
\tablehead{
\colhead{Parameter} & \colhead{Value} & \colhead{Units} \\
\colhead{(1)} & \colhead{(2)} & \colhead{(3)}
}
\startdata
$\beta_1$ & $0.545 \pm 0.018$ & -- \\
$r_{c1}$ & $4.9 \pm 1.1 $ & pixels \\
$e_1$ & $(1.00 \pm 0.70) \times 10^{-2}$ & -- \\
$\beta_2$ & $0.912 \pm 0.116$ & -- \\
$r_{c2}$ & $30.3 \pm 4.1$  & pixels \\
$e_2$ & $0.314 \pm 0.029$ & -- \\
$\theta$ & $2.912 \pm 0.016$ & radians \\
$f$ & $0.928 \pm 0.030$ & -- \\
$A_1$ & $(4.58 \pm 0.20) \times 10^{-7}$ & \si{cnt \,s^{-1}\,cm^{-2}} \\
$C$ & $(1.38 \pm 1.14) \times 10^{-10}$ & \si{cnt \,s^{-1}\,cm^{-2}} \\
\enddata
\tablecomments{Best-fit parameters for the 2D double-$\beta$ model. The model consists of inner (1) and outer (2) components with ellipticities $e$, position angle $\theta$, and flux fraction $f$, such that $A_2=A_1(1-f)$.}
\end{deluxetable}

\subsection{Morphological Parameters \label{subsec:morpho}}

As previously mentioned, the first indication of the dynamical regularity of SPT2215 arose from the evaluation of its X-ray morphological features. Here, we take advantage of our increased image statistics to compute traditional measures: concentration ($c$), centroid shift ($\omega$), and the third-order power ratio ($P_3/P_0$), as well as the morphology index ($\delta$) introduced by \cite{10.1093/mnras/staa2363}.  \\

The concentration parameter $c$ quantifies the surface brightness peak strength relative to the overall distribution and is mainly employed to identify the presence of cool cores. Here, we use the definition of $c$ as the ratio of the integrated X-ray fluxes under the core radius and the outer radius apertures. We choose $r_{\text{in}}=100$ kpc and $r_{\text{out}}=500$ kpc for consistency with \cite{2022MNRAS.513.3013Y}, therefore $c = {S_{100 \hspace{0.06cm} kpc}}/{S_{500 \hspace{0.06cm} kpc}}$.\\

From \cite{2006MNRAS.373..881P}, the centroid shift is defined as the standard deviation of the projected separation between the X-ray brightness peak and the model fitted center. These separations are computed along a series of $n=20$ decreasing apertures starting from $r_{out}= 500$ kpc. Thus, $\omega = \frac{1}{r_{out}}\left[\frac{1}{n-1} \sum_i (\Delta_i - \left<\Delta \right>)^2 \right]^{1/2}$, where $\Delta_i$ is the separation between the brightness peak and the fitted center in the ith aperture, and $\left<\Delta \right>$ is the mean separation value. Here, the center for each apperture is fitted with a single 2D $\beta$-model. This metric is sensitive to deviations from symmetry across multiple spatial scales, though for highly peaked surface brightness distributions, it can be dominated by the core contribution, as it is going to be further discussed in Section \ref{sec:4.3}. \\

Finishing the set of traditional morphological parameters, we compute the third-order power ratio $P_3/P_0$, which is particularly sensitive to substructures and departures from mirror symmetry \citep{2005ApJ...624..606J}. The powers are derived from the multipole expansion of the X-ray surface brightness within $r_{\text{out}}=500$ \citep{1995ApJ...452..522B,1996ApJ...458...27B}. \\

\begin{figure*}[ht!]
    \centering
    \includegraphics[width=\textwidth]{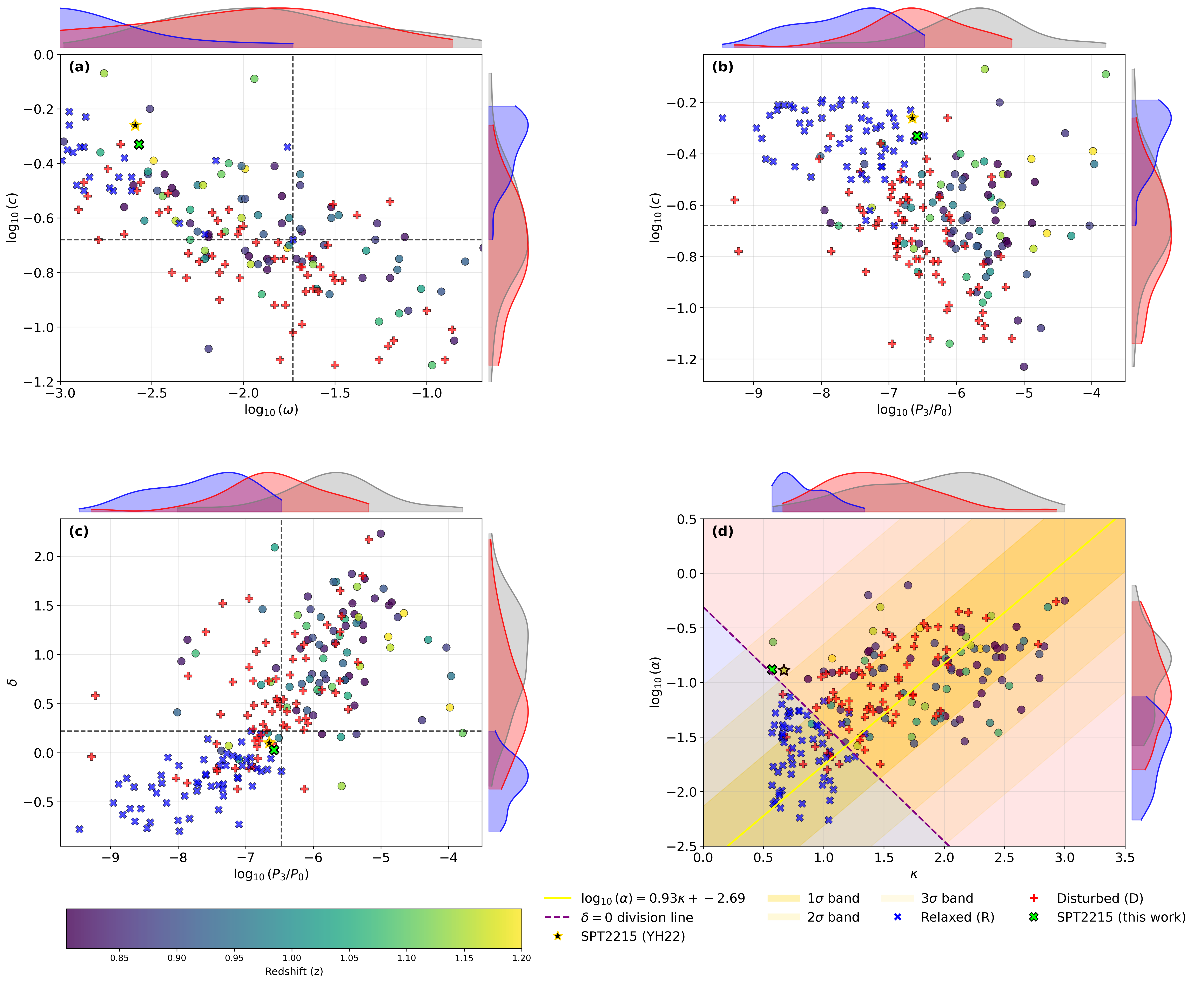}
    \caption{Parameter spaces of the morphological parameters considered in this work. Blue X’s and red crosses represent the relaxed and disturbed clusters from the test sample of \cite{10.1093/mnras/staa2363}, respectively, which have independent dynamical classifications. Circles indicate the selected high-redshift subsample ($0.8 < z < 1.2$) from \cite{2022MNRAS.513.3013Y}. The blue, red, and grey shaded regions on the outside of the plotting frames illustrate the distribution of relaxed, disturbed, and high-redshift clusters, respectively, in the parameter spaces. Dashed black lines mark the threshold values limiting the region occupied by relaxed clusters.}
    \label{fig:morpho}
\end{figure*}

The morphology index $\delta$ for SPT2215 was previosly calculated in \cite{2022MNRAS.513.3013Y} using one single observation (obsID$=24614$), giving $\delta=0.10\pm0.01$. Here, we check if this result holds for our merged image.  The morphology index is defined as the distance to the line that best separates the dynamical state of a sample of 115 well studied clusters in the $\kappa-\alpha$ space: $\delta = 0.68\log_{10}(\alpha) + 0.73\kappa + 0.21$. Here, the assymetric factor $\alpha$ and the profile parameter $\kappa$ are independent measures given by $\alpha = {\sum\limits_{x,y} \left[ s(x,y) - s(x',y') \right]^2}/{\sum\limits_{x,y} s^2(x,y)}$, where $s(x',y')$ is the flux at the symmetry pixel of $(x,y)$ with respect to the center, and $\kappa = ({1+e})/{\beta}$, where $e$ and $\beta$ have the same meaning as before and are extracted from a single elliptical $\beta$-model fit. To ensure consistency with the empirical formula for $\delta$, we adopt the same aperture and smoothing parameters used in \cite{10.1093/mnras/staa2363} when computing $\alpha$ and $\kappa$. We obtain $\log_{10}{(c)} = -0.33$, $\log_{10}{\omega} = -2.57$, $\log_{10}(P_3/P_0) = -6.58$, $\log_{10}{(\alpha)} = -0.88$, $\kappa = 0.57$, and $\delta = 0.028$. These values are in good agreement to those reported in \cite{2022MNRAS.513.3013Y}, as showed in Figure \ref{fig:morpho}, suggesting that their convergence is not strongly limited by data quality or exposure time. We further discuss in Section \ref{sec:4.3} the numerical correlations among the considered metrics, as well as potential decouplings from the underlying physical phenomenology they are meant to trace. 

\section{Discussion} \label{sec:discu}

\subsection{Core Sloshing} \label{sec:4.1}

The thermodynamic characterization of SPT2215's ICM presented in this work reveals notable anisotropies, including what appears to be contact discontinuities along the SE–NW direction, approximately located at the estimated cooling radius,  $r_{cool}=147\pm10$ kpc. The observed sharp increase in temperature and entropy within the corresponding radial bin, accompanied by an approximately constant pressure, are well-established signatures of cold fronts. In Figure \ref{fig:dirpro}, the SE-NW temperature and entropy profiles are consistent with a core oscillation within the boundaries of $r_{\text{cool}}$, where, below $r_{cool}$, the gas in the SE is considerably colder compared to other directions. This mismatch then inverts just beyond $r_{cool}$, with the entropy finally converging in the $210$-$410$ kpc bin. This radial behaviour aligns with the central emission excess distribution seen in the residual map of Figure \ref{fig:residual}, which shows clear spiral patterns along multiple scales. \\

%From the azimuthally-averaged pressure profile, we probe the motion of the cold gas similarly to \cite{2001ApJ...551..160V}. The bin correspondent to the stagnation point is at $r_0\approx137\pm18$, while the free stream bin is at $r_1\approx247\pm30$. Since min($P_0/P_1$)$=1.5$ We estimate a lower bound Mach number of $\mathcal{M_{\text{lower}}}=0.73$, corresponding to a velocity of $v\approx1140$ km/s. This is way above the expected velocity for sloshing motions \\

Spatial resolution limits the detection of cold fronts in clusters as distant as SPT2215. Even with the combined observations analyzed in this work, we do not detect them solely from the surface brightness. Nevertheless, the scenario supported by our spectral modeling is not unexpected, as many cool-core clusters contain signatures of cold fronts. The hydrodynamic simulations of mergers done by \cite{2006ApJ...650..102A} revealed that core sloshing can induce such discontinuities given a sufficiently steep entropy profile. When fitting the azimuthaly-averaged entropy profile with a power law, we get a slope of $1.12\pm0.12$, as predicted by self-similar evolution models \citep{2001ApJ...546...63T,2005MNRAS.364..909V,2009ApJS..182...12C}. Restricting the fit for the [$0.01$-$0.08$]$r_{180}$ range, we find a slope of $1.25\pm0.18$. The mean scaled entropy profile slope of cold front clusters reported by \cite{2010A&A...516A..32G} is $\alpha[(0.01\text{-}0.08)r_{180}]=1.22\pm0.01$. Although our central value is consistent with theirs, the uncertainty is large enough that we cannot distinguish this slope from the self-similar expectation at a statistically significant level. If the entropy profile is indeed steeper in this radial range, a past off-axis interaction with a gas-poor substructure could have significantly perturbed the low-entropy gas in the cluster’s core, giving rise to sloshing cold fronts that propagated outward with time, stalling approximately at the cooling radius. \\

It is worth noting the coincident SE-NW orientation of the extended blue filaments in the BCG, as observed by archival HST/WFC3-UVIS F200LP imaging (Program ID: idgbt2020). Magnetohydrodynamic simulations indicate that sloshing-driven motions can amplify magnetic field strengths along shear flows, suppress gas mixing, and promote anisotropic thermal conduction \citep{2011ApJ...743...16Z,2013ApJ...762...69Z}, which could be a plausible mechanism for the apparent SE–NW orientation of the filaments.
%It is plausible that sloshing-driven shear motions, aligned with draped magnetic fields, enable anisotropic thermal conduction that creates a preferential channel for the inflow cool of gas. 
A minimum local cooling time over the free-fall time $t_{\text{cool}}/t_{\text{ff}}$ below $\sim10$ indicate conditions favorable for thermal instability and efficient multiphase condensation in the circumgalactic medium \citep{2015Natur.519..203V}. We find a min$(t_{\text{cool}}/t_{\text{ff}})\approx8$ at  $r\approx25$ kpc, indicating that the resulting condensation is fueling the intense starburst in the BCG \citep{2023ApJ...947...44C}.  \\

%Figure \ref{fig:dirpro} shows that the temperature remains nearly constant along the SW–NE axis out to $r\sim180$ kpc. The geometry of the opposed dipolar discontinuities seen in the temperature map of Figure \ref{fig:thermo_maps} is consistent with cold fronts viewed close to the plane of the sky. This makes it unlikely that the observed flattening of the central temperature profile is due to projection effects.  We suggest that bulk motions associated with sloshing are redistributing the low-entropy gas to larger scales, thereby flattening the temperature gradient across the core. We use the Brunt–Väsälä frequency $\omega_{\text{BV}}$ \citep{1980tsp..book.....C} to calculate the sloshing period $P=2\pi/\omega_{\text{BV}}$ and estimate the sloshing time-scale to be $P/2=$ ... \\

\subsection{Ghost Bubble Candidate} \label{sec:4.2}

Let us now discuss the plausibility of the possible X-ray cavity revealed by our imaging analysis. The majority of cool core clusters harbour radio sources and bubble-like structures, filled with relativistic electrons, produced by AGN jets, likely triggered by accretion of cold gas onto the central supermassive black hole. The propagation of these bubbles is an efficient way of injecting energy into the central regions of the cluster, since it creates pressure waves that are viscously dissipated throughout the ICM. The ASKAP weak radio detection at 0.9 GHz surrounding SPT2215's BCG indicates ongoing AGN feedback. We do not expect GHz emission from a bubble with such high age ($t\approx190$ Myr), making it a ghost bubble candidate. Considering the observed cavity properties, we calculate a jet power that is consistent with the value obtained from the $P_{\text{cav}}$-$L_{\text{cool}}$ relation \citep{2010ApJ...720.1066C} reported by \cite{2023ApJ...947...44C}, but with a significant uncertainty improvement: $\log_{10}(P_{\text{cav}}/10^{42} \text{erg s}^{-1})=2.66\pm0.23$ against $\log_{10}(P_{\text{cav}}/10^{42} \text{erg s}^{-1})=2.6\pm 0.3 \text{ (stat)} \pm 0.8 \text{ (int. scatter)}$. Naturally, there are hard limits on the cavity's measurable features conditioned to spatial resolution and growth limit. In our case, the cavity's dimensions exceed the detection threshold for the considered redshift, while the high mean temperature of the surrounding ICM places it below the expected growth limit. From a sample of 148 cavities present in 69 targets, \cite{2016ApJS..227...31S} found a linear correlation between the cavity area and its projected distance expressed by $\log_{10} (A)=(1.94\pm0.04)\log_{10} (D) + 0.02 \pm 0.05$. Given $A\approx 4700\pm1400$ kpc$^2$ and $D\approx230$ kpc, we find a deviation of $\Delta \log_{10}(A) \approx -0.93$ from the considered correlation, which corresponds to nearly $3\sigma$. For this projected distance, the cavity's size should be about an order of magnitude larger. This could be seen as a main caveat of this observed feature as a real structure. On the other hand, it is worth to consider that the sample of clusters used by \cite{2016ApJS..227...31S} did not include any ghost bubbles. The growth conditions in high-redshift clusters can differ significantly due to evolving ICM properties (e.g., magnetic pressure, viscosity, thermal conduction), which may affect the morphology of highly evolved bubbles and cause deviations from general scaling relations (see \citealt{2023Galax..11...73B} for a review).

\subsection{Cooling-Feedback Cycle}

Strong evidence supports the so called cooling–feedback model, in which radiative cooling and feedback from the central supermassive black hole mutually influence and constrain each other in a self-regulating manner (eg., \citealt{2006MNRAS.373..959D,2017MNRAS.464.4360M,2024ApJ...976..169C}). Following the feedback-regulated precipitation framework developed by \citet{2017ApJ...845...80V} for cluster cores, we can attempt to qualitatively assess the current phase of SPT2215. In order to so, we consider three independent probes: The time since the AGN restarted, the minimum value of $t_{\text{cool}}/t_{\text{ff}}$ and the star formation rate in the BCG.
If we take the cavity age calculated in Section \ref{sec:resid} ($t_{\text{cav}}\approx190\pm20$ Myr) as a lower limit for the time since the last outburst, then SPT2215 would be likely in the early phase of progression towards a long-lasting self-regulated cycle. The found min$(t_{\text{cool}}/t_{\text{ff}})\approx8$ is under the threshold for thermal instability, indicating that the isentropic region out to $r\approx25$ kpc is still not completely formed. Considering that a steep central entropy gradient can only develop following the formation of a stable hot halo, a process expected to occur no earlier than $z \sim 2\text{-}2.5$ for massive systems (see \citealt{2016A&ARv..24...14O}), a cluster observed at $z = 1.16$ would be expected to have at most roughly $2$ Gyr to have established a stable cooling-feedback loop. Therefore, it is reasonable to conclude that SPT2215 is, at most, about to enter the second long-lasting self-regulated cycle. The idealized simulation of a cluster core done by \cite{2015ApJ...811...73L} (see Figure 2 of their paper) showed that the onset of star formation in the BCG can occur rapidly after the outburst event, leading to a drastic increase in the local buildup of cold gas. On the other hand, the response of min$(t_{\text{cool}}/t_\text{ff})$ to the outflows happens in a considerably longer timescale. This delay arises from the fact that, before a stable isentropic core can develop, the bulk low-entropy gas needs to be removed or reprocessed via turbulence, mixing and condensation. $\sim190$ Myr after the second outburst, the star formation rate would have enough time to be already stabilized around its peak, explaining the massive starburst in the BCG (SFR$=320^{+230}_{-140} M_{\odot} \text{yr}^{-1}$) \citep{2023ApJ...947...44C}. The thermodynamic profiles along the supposed bipolar outflow axis (SW–NE) show notable deviations compared to the SE–NW direction, including a central temperature bump in the NE direction (Figure \ref{fig:dirpro}) and entropy flattening at intermediate scales (Figure \ref{fig:radpro}). These asymmetries likely reflect enhanced inhomogeneities and turbulence induced by the recent AGN activity. In particular, uplift and mixing along the outflow axis can lead to transient entropy irregularities and deviations from hydrostatic equilibrium. As the system evolves toward a fully precipitation-regulated state, buoyancy and turbulent damping are expected to progressively erase these features, leading to a more symmetric and smoother thermodynamic structure.

\subsection{Limitations of the Morphology-based Classification} \label{sec:4.3}

The use of morphological parameters as proxies for cluster relaxation is based on the principle that the 2D projection of the emission reliably traces dynamical phenomena. Given X-ray observables and the underlying intrinsic state, the morphological measures serve as estimators of the true dynamical state. The mapping is non-injective, that is, distinct intrinsic states may produce similar 2D features as a result of projection or fundamentally different physical conditions, which is why classifications based on single measures must be taken with caution. Instead, the set of morphological parameters usually populate a multidimensional space where certain regions are empirically associated to relaxed and perturbed clusters. Thresholds are typically determined using a sample of clusters with independently confirmed dynamical states, or derived numerically from mock images of simulated clusters. \\

Let us now explore potential caveats of the presented framework and how it relates to the current classification of SPT2215. First, thresholds are naturally set to maximize the purity of relaxed/perturbed subsamples, which inevitable tends to misclassify systems with intermediate dynamical stages. This effect is likely to be more prevalent for high-redshift clusters due to the expected higher merger rate. Second, a fixed threshold assumes either that (1) the correlation between the 2D observable and a given 3D property associated to the dynamical state does not have a significant mass- or redshift-dependent evolution, or (2) the functional form of this evolution is known a priori, such that the parameter can be appropriately rescaled. This matter was addressed by \cite{2021MNRAS.503.3394C} (hereafter CBV21) using Mock-X \citep{2021MNRAS.506.2533B} to generate synthetic X-ray images of clusters from the IllustrisTNG, BAHAMAS and MACSIS cosmological hydrodynamical simulations. They evaluated various aspects of the established morphological measures, including how they are correlated to each other and to 3D theoretical parameters extracted from the simulated halos. Regarding redshift evolution, although many specific considerations can be made about the impact of subgrid physics and numerical resolution on the observed trends for each simulation, CBV21 found that generally all morphological parameters exhibit some level of redshift evolution. This result highlights the critical importance of correctly modeling the functional form of this dependence. \\

Figure \ref{fig:morpho} shows the distribution of clusters in the relevant parameter spaces for three subsamples: relaxed ($R$), disturbed ($D$) and high-redshift ($0.8<z<1.2$) clusters. We may give special attention to parameters that are somehow dependent on the cluster's central emission cuspiness: more readily, the concentration index ($c$), and indirectly, the centroid shift ($\omega$). First, we notice from Figure \ref{fig:morpho} that, with the exception of $c$ and $\omega$, all other parameters tend to produce a displaced distribution towards the unrelaxed end for the high-redshift subsample. Additionally, $c\times\omega$ gives the highest contamination for a relaxed region among all parameter spaces. We shall briefly discuss the origin of this trend. To begin, we must acknowledge a degree of numerical degeneracy between $c$ and $\omega$, since a centrally peaked brightness distribution can supress the measured centroid displacements across all apertures, weighting the centroid more strongly toward the core. Naturally, if one has to use a single morphological parameter to classify the dynamical state of a test sample, this feature can be seen as a strength, since it somehow captures two different aspects of the emission distribution: the central cuspiness, indicating the presence of a structured cool core, and assymetries, quantifying the presence of substructures. For large samples, it can be effective in distinguishing highly dynamically disturbed clusters from very relaxed ones. For instance, \cite{2008ApJS..174..117M} used a core-excised $\omega$ as a single relaxation proxy for a sample of 115 clusters in the range $0.1<z<1.3$, finding a clear absence of highly relaxed clusters at $z>0.5$. We argue that this degeneracy becomes particularly problematic for high-redshift clusters when the core contribution is included. We suggest that the low purity for $c\times\omega$ seen in Figure \ref{fig:morpho}(a) is caused by a combination of this degeneracy and a poorly modeled redshift dependence of $c$. \\

The surface brightness peakiness ($p$) from SPA is a more refined version of $c$, since it is measured under a same scaled surface brightness level by assuming a self-similar model, and it includes a $(1+z)$ term, which results in a nearly constant fraction of peaked clusters across redshift. The other two statistics are the symmetry ($s$) and the alingment ($a$) as defined in \cite{2015MNRAS.449..199M}. For SPT2215, the SPA measurements are $s=1.24\pm0.15$, $p=-0.46\pm0.04$ and $a=1.42\pm0.16$ \citep{2023ApJ...947...44C}. Upon visual inspection of Figure 2 of their paper, one can notice that the $a\times s$ space yields a higher contamination compared to $s \times p$ and $a \times p$, suggesting that $p$ carries a greater wheight in distinguishing the dynamical state. CBV21 points out that the redshift evolution of $p$, even with the $(1+z)$ term, is too strong, producing larger subsamples of relaxed clusters with increasing redshift. By fitting the morphological parameters with a $(1+z)^{-\beta}$ factor, they found $\beta(s)=0.59\pm0.18$, $\beta(p)=3.42\pm0.49$ and $\beta(a)=1.38\pm0.21$. Applying these corrections to the original SPA measurements, we get $s_z=1.04\pm0.16$, $p_z=-1.60\pm0.17$ and $a_z=0.96\pm0.17$. For the sake of comparison, if we keep the original relaxation thresholds ($s>0.87$, $p>-0.82$ and $a>1.00$), then $s$ marginally passes the criterion, $p$ strongly fails and $a$ remains in the borderline. Therefore, once the redshift evolution is accounted for, we find a scenario that is much more consistent with an intermediate dynamical state, which matches the near-zero morphology index $\delta=0.028$ and the more central $\log_{10}(P_3/P_0)=-6.58$ value (although more consistent with the disturbed subsample distribution in this case). \\

\subsection{Physical Interpretation of Cuspiness Enhancement}

It seems that, at high-redshift, a centrally peaked emission is not as strong an indicator of a virialized state as it is in the local universe. The most straightforward reason would be that a
surviving post-merger cool core under a high merger rate
regime, can retain a high peakiness, even when the cluster remains
dynamically disturbed. But we argue this can also be the contribution of competing physical processes, particularly related to a not yet fully regulated AGN feedback cycle at earlier cosmic times, as pointed by \cite{2024ApJ...976..169C}. Our combined analysis, together with the interpretation given in Section \ref{sec:4.3}, supports the idea that SPT2215 is experiencing extreme thermal imbalance during the period between an outburst event and the development of a two-phase entropy structure, which can be creating a cooling-dominated regime that amplifies the surface brightness cuspiness while not necessarily reflecting strong virialization. Thus, what we might be seeing in SPT2215 is a short transient state in which feedback has been recently triggered and has not yet had sufficient time to offset the overwhelming cooling, which in turn dominates the central X-ray emission and bias morphological criteria towards relaxation.

\section{Summary} \label{sec:5}

We present a detailed revision of the dynamical state of the high-redshift galaxy cluster SPT-CL J2215-3537, based on a comprehensive X-ray analysis of all currently available Chandra observations, and provide a discussion on the use of morphological statistics to assess relaxation in the high-redshift regime. Our results and conclusions can be summarized as follows: 

\begin{enumerate}
    \item The thermodynamic characterization of SPT2215 reveals significant anisotropies, including off-center clumps of colder gas toward the eastern direction and features consistent with cold front signatures near the cooling radius along the SE–NW axis. The $\beta$-model-subtracted residual show a central dipolar pattern along the SE–NW direction and arc-shaped surface brightness excess across different scales, both features consistent with a core sloshing scenario. %The axis of the opposite cold fronts is coincident with the orientation of the extendend blue filaments in the BCG, as shown by the HST/F200LP observation.
    \item We propose that a past off-axis merger with a gas-poor structure triggered core sloshing. The steep entropy profile in the core would allow the resulting oscillatory, subsonic gas motions to generate the observed cold fronts near the edges of the cool core. These bulk motions may also create a preferential channel for the inflow of low-entropy gas, potentially enhancing star formation along the SE–NW axis. Even if the cluster does not qualify as the most distant relaxed cluster, as claimed by previous works, to our knowledge it is the most distant cluster where sloshing cold fronts were observed.
    \item The residual map reveals a candidate ghost bubble located approximately $230$ kpc to the NW. Although the observed features of the cavity deviate significantly from known empirical relations, the derived $P_{\text{cav}}$ is consistent with the previous estimative from the $P_{\text{cav}}$-$L_{\text{cool}}$ scaling relation, indicating that radiative cooling is currently outweighing AGN feedback.
    \item The main properties of SPT2215's core can be qualitatively explained by the precipitation-regulated framework developed by \cite{2015Natur.519..203V}. Our combined analysis indicates that SPT2215 is currently in the early phase of progression towards a long-lasting self-regulated cooling-feedback cycle. At present, condensation appears to be mostly confined to the inner regions ($\sim25\pm6$ kpc) and the entropy distribution at each radius is still inhomogeneous out to $\sim180\pm40$ kpc along the SW-NE axis.
    \item We compute traditional morphological parameters for our merged X-ray image and find a morphology index $\delta=0.028$ and power ratio $\log_{10}(P_3/P_0)=-6.58$ consistent with a mildly disturbed system, but a concentration $\log_{10}(c)=-0.33$ and centroid shift $\log_{10}(\omega)=-2.57$ characteristic of highly relaxed systems. We discuss evidence suggesting that these values are biased due to a strong redshift-evolution of morphological parameters associated to the central emission cuspiness. We reevaluate the SPA criteria, which had previously classified SPT2215 as dynamically relaxed, noticing that the peakiness statistic carries a greater weight in placing the system in the relaxed locus of the parameter space. When applying the redshift dependences indicated by CBV21, we find new SPA measurements ($s_z=1.04\pm0.16$, $p_z=-1.60\pm0.17$ and $a_z=0.96\pm0.17$), which are now consistent with an intermediate dynamical state. 
    \item We propose that the strong redshift-dependence of the peakiness and concentration parameters is related to an undeveloped cooling-feedback cycle at high-redshift. In this sense, a peaked cluster at high-redshift would not imply a virialized dynamical state as strongly as it does at low redshift due to the increased prevalence of transient or rapidly evolving core conditions at earlier cosmic times. Spatially resolved X-ray analysis of other similar and, especially, higher redshift clusters will be crucial to further test this hypothesis (Bessa et al., in prep).
  
\end{enumerate}

 Our joint analysis indicates that SPT2215 is a mildly disturbed high-redshift cluster hosting a cool core. We reiterate the relevance of follow-up observations of SPT2215, since this system represents an ideal case for understanding the cooling-feedback cycle and out-of-equilibrium conditions in the cores of high-redshift galaxy clusters. Observations with the future Advanced X-ray Imaging Satellite (AXIS) \citep{2024Univ...10..273R} will be particularly decisive, allowing for enough statistics to generate thermodynamic maps with significantly enhanced spatial resolution and coverage, probing potential new features such as gas motions beyond the cold fronts, abundance distributions, fainter X-ray cavities and more.  Meanwhile, additional independent constraints on clusters dynamical state will be particularly important, such as those, that can be provided through measurements of $f_{\text{ICL}}$ in a set of bands that best probe the intracluster light excess, corresponding to the rest-frame wavelength interval $5200$–$7300$ $\textup{~\AA}$ \citep{2024ApJ...960L...7J}.

\begin{acknowledgments}
We are deeply grateful to Dr.~Alberto Ardila and Vinicius Sanches for the insightful discussions. This work was supported by the \textit{Coordenação de Aperfeiçoamento de Pessoal de Nível Superior} (CAPES) and the \textit{Fundação de Amparo à Pesquisa do Estado do Rio de Janeiro} (FAPERJ). R.A.D. acknowledges partial support from CNPq grant 312565/2022-4. Y.J-T. acknowledges financial support from the State Agency for Research of the Spanish MCIU through Center of Excellence Severo Ochoa award to the Instituto de Astrofísica de Andalucía CEX2021-001131-S funded by MCIN/AEI/10.13039/501100011033, and from the grant PID2022-136598NB-C32 Estallidos and project ref. AST22-00001-Subp-15 funded by the EU-NextGenerationEU. This research employs a list of Chandra datasets, obtained by the Chandra X-ray Observatory, contained in~\dataset[DOI: 10.25574/cdc.504]{https://doi.org/10.25574/cdc.504}, and the software provided by the High Energy Astrophysics Science Archive Research Center (HEASARC), which is a service of the Astrophysics Science Division at NASA/GSFC. We also used Ned Wright's CosmoCalc \citep{ned}.

An independent study of SPT2215 using the same Chandra dataset analyzed in this work was published by \cite{2026ApJ...997..320S}. 

\end{acknowledgments}

\bibliography{sample631}{}

@ARTICLE{2006MNRAS.368..497D,
       author = {{Diehl}, Steven and {Statler}, Thomas S.},
        title = "{Adaptive binning of X-ray data with weighted Voronoi tessellations}",
      journal = {\mnras},
         year = 2006,
        month = may,
       volume = {368},
       number = {2},
        pages = {497-510},
          doi = {10.1111/j.1365-2966.2006.10125.x},
archivePrefix = {arXiv},
       eprint = {astro-ph/0512074},
 primaryClass = {astro-ph},
       adsurl = {https://ui.adsabs.harvard.edu/abs/2006MNRAS.368..497D}
}

@ARTICLE{2003MNRAS.342..345C,
       author = {{Cappellari}, Michele and {Copin}, Yannick},
        title = "{Adaptive spatial binning of integral-field spectroscopic data using Voronoi tessellations}",
      journal = {\mnras},
         year = 2003,
        month = jun,
       volume = {342},
       number = {2},
        pages = {345-354},
          doi = {10.1046/j.1365-8711.2003.06541.x},
archivePrefix = {arXiv},
       eprint = {astro-ph/0302262},
 primaryClass = {astro-ph},
       adsurl = {https://ui.adsabs.harvard.edu/abs/2003MNRAS.342..345C}
}

@ARTICLE{2023A&A...676A..39J,
       author = {{Jim{\'e}nez-Teja}, Y. and {Dupke}, R.~A. and {Lopes}, P.~A.~A. and {V{\'\i}lchez}, J.~M.},
        title = "{Dissecting the RELICS cluster SPT-CLJ0615-5746 through intracluster light: Confirmation of the multiple merging state of the cluster formation}",
      journal = {\aap},
         year = 2023,
        month = aug,
       volume = {676},
          eid = {A39},
        pages = {A39},
          doi = {10.1051/0004-6361/202346580},
archivePrefix = {arXiv},
       eprint = {2305.10860},
 primaryClass = {astro-ph.GA},
       adsurl = {https://ui.adsabs.harvard.edu/abs/2023A&A...676A..39J}
}

@ARTICLE{2015MNRAS.449..199M,
       author = {{Mantz}, Adam B. and {Allen}, Steven W. and {Morris}, R. Glenn and {Schmidt}, Robert W. and {von der Linden}, Anja and {Urban}, Ondrej},
        title = "{Cosmology and astrophysics from relaxed galaxy clusters - I. Sample selection}",
      journal = {\mnras},
         year = 2015,
        month = may,
       volume = {449},
       number = {1},
        pages = {199-219},
          doi = {10.1093/mnras/stv219},
archivePrefix = {arXiv},
       eprint = {1502.06020},
 primaryClass = {astro-ph.CO},
       adsurl = {https://ui.adsabs.harvard.edu/abs/2015MNRAS.449..199M}
}

@ARTICLE{2020MNRAS.495..705Z,
       author = {{Zenteno}, A. and {Hern{\'a}ndez-Lang}, D. and {Klein}, M. and {Vergara Cervantes}, C. and {Hollowood}, D.~L. and {Bhargava}, S. and {Palmese}, A. and {Strazzullo}, V. and {Romer}, A.~K. and {Mohr}, J.~J. and {Jeltema}, T. and {Saro}, A. and {Lidman}, C. and {Gruen}, D. and {Ojeda}, V. and {Katzenberger}, A. and {Aguena}, M. and {Allam}, S. and {Avila}, S. and {Bayliss}, M. and {Bertin}, E. and {Brooks}, D. and {Buckley-Geer}, E. and {Burke}, D.~L. and {Capasso}, R. and {Carnero Rosell}, A. and {Carrasco Kind}, M. and {Carretero}, J. and {Castander}, F.~J. and {Costanzi}, M. and {da Costa}, L.~N. and {De Vicente}, J. and {Desai}, S. and {Diehl}, H.~T. and {Doel}, P. and {Eifler}, T.~F. and {Evrard}, A.~E. and {Flaugher}, B. and {Floyd}, B. and {Fosalba}, P. and {Frieman}, J. and {Garc{\'\i}a-Bellido}, J. and {Gerdes}, D.~W. and {Gonzalez}, J.~R. and {Gruendl}, R.~A. and {Gschwend}, J. and {Gutierrez}, G. and {Hartley}, W.~G. and {Hinton}, S.~R. and {Honscheid}, K. and {James}, D.~J. and {Kuehn}, K. and {Lahav}, O. and {Lima}, M. and {McDonald}, M. and {Maia}, M.~A.~G. and {March}, M. and {Melchior}, P. and {Menanteau}, F. and {Miquel}, R. and {Ogando}, R.~L.~C. and {Paz-Chinch{\'o}n}, F. and {Plazas}, A.~A. and {Roodman}, A. and {Rykoff}, E.~S. and {Sanchez}, E. and {Scarpine}, V. and {Schubnell}, M. and {Serrano}, S. and {Sevilla-Noarbe}, I. and {Smith}, M. and {Soares-Santos}, M. and {Suchyta}, E. and {Swanson}, M.~E.~C. and {Tarle}, G. and {Thomas}, D. and {Varga}, T.~N. and {Walker}, A.~R. and {Wilkinson}, R.~D. and {DES Collaboration}},
        title = "{A joint SZ-X-ray-optical analysis of the dynamical state of 288 massive galaxy clusters}",
      journal = {\mnras},
         year = 2020,
        month = jun,
       volume = {495},
       number = {1},
        pages = {705-725},
          doi = {10.1093/mnras/staa1157},
archivePrefix = {arXiv},
       eprint = {2004.01721},
 primaryClass = {astro-ph.GA},
       adsurl = {https://ui.adsabs.harvard.edu/abs/2020MNRAS.495..705Z}
}

@ARTICLE{2015A&A...575A.127P,
       author = {{Parekh}, Viral and {van der Heyden}, Kurt and {Ferrari}, Chiara and {Angus}, Garry and {Holwerda}, Benne},
        title = "{Morphology parameters: substructure identification in X-ray galaxy clusters}",
      journal = {\aap},
         year = 2015,
        month = mar,
       volume = {575},
          eid = {A127},
        pages = {A127},
          doi = {10.1051/0004-6361/201424123},
archivePrefix = {arXiv},
       eprint = {1411.6525},
 primaryClass = {astro-ph.CO},
       adsurl = {https://ui.adsabs.harvard.edu/abs/2015A&A...575A.127P}
}

@ARTICLE{2007A&A...467..485H,
       author = {{Hashimoto}, Yasuhiro and {B{\"o}hringer}, H. and {Henry}, J.~P. and {Hasinger}, G. and {Szokoly}, G.},
        title = "{Robust quantitative measures of cluster X-ray morphology, and comparisons between cluster characteristics}",
      journal = {\aap},
         year = 2007,
        month = may,
       volume = {467},
       number = {2},
        pages = {485-499},
          doi = {10.1051/0004-6361:20065125},
archivePrefix = {arXiv},
       eprint = {astro-ph/0611804},
 primaryClass = {astro-ph},
       adsurl = {https://ui.adsabs.harvard.edu/abs/2007A&A...467..485H}
}

@ARTICLE{2008MNRAS.383..879A,
       author = {{Allen}, S.~W. and {Rapetti}, D.~A. and {Schmidt}, R.~W. and {Ebeling}, H. and {Morris}, R.~G. and {Fabian}, A.~C.},
        title = "{Improved constraints on dark energy from Chandra X-ray observations of the largest relaxed galaxy clusters}",
      journal = {\mnras},
         year = 2008,
        month = jan,
       volume = {383},
       number = {3},
        pages = {879-896},
          doi = {10.1111/j.1365-2966.2007.12610.x},
archivePrefix = {arXiv},
       eprint = {0706.0033},
 primaryClass = {astro-ph},
       adsurl = {https://ui.adsabs.harvard.edu/abs/2008MNRAS.383..879A}
}

@ARTICLE{2004MNRAS.353..457A,
       author = {{Allen}, S.~W. and {Schmidt}, R.~W. and {Ebeling}, H. and {Fabian}, A.~C. and {van Speybroeck}, L.},
        title = "{Constraints on dark energy from Chandra observations of the largest relaxed galaxy clusters}",
      journal = {\mnras},
         year = 2004,
        month = sep,
       volume = {353},
       number = {2},
        pages = {457-467},
          doi = {10.1111/j.1365-2966.2004.08080.x},
archivePrefix = {arXiv},
       eprint = {astro-ph/0405340},
 primaryClass = {astro-ph},
       adsurl = {https://ui.adsabs.harvard.edu/abs/2004MNRAS.353..457A}
}

@ARTICLE{2014ApJ...792...25N,
       author = {{Nelson}, Kaylea and {Lau}, Erwin T. and {Nagai}, Daisuke},
        title = "{Hydrodynamic Simulation of Non-thermal Pressure Profiles of Galaxy Clusters}",
      journal = {\apj},
         year = 2014,
        month = sep,
       volume = {792},
       number = {1},
          eid = {25},
        pages = {25},
          doi = {10.1088/0004-637X/792/1/25},
archivePrefix = {arXiv},
       eprint = {1404.4636},
 primaryClass = {astro-ph.CO},
       adsurl = {https://ui.adsabs.harvard.edu/abs/2014ApJ...792...25N}
}

@ARTICLE{2014MNRAS.443.1973V,
       author = {{von der Linden}, Anja and {Mantz}, Adam and {Allen}, Steven W. and {Applegate}, Douglas E. and {Kelly}, Patrick L. and {Morris}, R. Glenn and {Wright}, Adam and {Allen}, Mark T. and {Burchat}, Patricia R. and {Burke}, David L. and {Donovan}, David and {Ebeling}, Harald},
        title = "{Robust weak-lensing mass calibration of Planck galaxy clusters}",
      journal = {\mnras},
         year = 2014,
        month = sep,
       volume = {443},
       number = {3},
        pages = {1973-1978},
          doi = {10.1093/mnras/stu1423},
archivePrefix = {arXiv},
       eprint = {1402.2670},
 primaryClass = {astro-ph.CO},
       adsurl = {https://ui.adsabs.harvard.edu/abs/2014MNRAS.443.1973V}
}

@ARTICLE{2012ApJ...751..121N,
       author = {{Nelson}, Kaylea and {Rudd}, Douglas H. and {Shaw}, Laurie and {Nagai}, Daisuke},
        title = "{Evolution of the Merger-induced Hydrostatic Mass Bias in Galaxy Clusters}",
      journal = {\apj},
         year = 2012,
        month = jun,
       volume = {751},
       number = {2},
          eid = {121},
        pages = {121},
          doi = {10.1088/0004-637X/751/2/121},
archivePrefix = {arXiv},
       eprint = {1112.3659},
 primaryClass = {astro-ph.CO},
       adsurl = {https://ui.adsabs.harvard.edu/abs/2012ApJ...751..121N}
}

@ARTICLE{2004MNRAS.351..237R,
       author = {{Rasia}, Elena and {Tormen}, Giuseppe and {Moscardini}, Lauro},
        title = "{A dynamical model for the distribution of dark matter and gas in galaxy clusters}",
      journal = {\mnras},
         year = 2004,
        month = jun,
       volume = {351},
       number = {1},
        pages = {237-252},
          doi = {10.1111/j.1365-2966.2004.07775.x},
archivePrefix = {arXiv},
       eprint = {astro-ph/0309405},
 primaryClass = {astro-ph},
       adsurl = {https://ui.adsabs.harvard.edu/abs/2004MNRAS.351..237R}
}

@ARTICLE{2006MNRAS.369.2013R,
       author = {{Rasia}, E. and {Ettori}, S. and {Moscardini}, L. and {Mazzotta}, P. and {Borgani}, S. and {Dolag}, K. and {Tormen}, G. and {Cheng}, L.~M. and {Diaferio}, A.},
        title = "{Systematics in the X-ray cluster mass estimators}",
      journal = {\mnras},
         year = 2006,
        month = jul,
       volume = {369},
       number = {4},
        pages = {2013-2024},
          doi = {10.1111/j.1365-2966.2006.10466.x},
archivePrefix = {arXiv},
       eprint = {astro-ph/0602434},
 primaryClass = {astro-ph},
       adsurl = {https://ui.adsabs.harvard.edu/abs/2006MNRAS.369.2013R}
}

@ARTICLE{1996ApJ...469..494E,
       author = {{Evrard}, August E. and {Metzler}, Christopher A. and {Navarro}, Julio F.},
        title = "{Mass Estimates of X-Ray Clusters}",
      journal = {\apj},
         year = 1996,
        month = oct,
       volume = {469},
        pages = {494},
          doi = {10.1086/177798},
archivePrefix = {arXiv},
       eprint = {astro-ph/9510058},
 primaryClass = {astro-ph},
       adsurl = {https://ui.adsabs.harvard.edu/abs/1996ApJ...469..494E}
}

@ARTICLE{2008MNRAS.384.1567M,
       author = {{Mahdavi}, A. and {Hoekstra}, H. and {Babul}, A. and {Henry}, J.~P.},
        title = "{Evidence for non-hydrostatic gas from the cluster X-ray to lensing mass ratio}",
      journal = {\mnras},
         year = 2008,
        month = mar,
       volume = {384},
       number = {4},
        pages = {1567-1574},
          doi = {10.1111/j.1365-2966.2007.12796.x},
archivePrefix = {arXiv},
       eprint = {0710.4132},
 primaryClass = {astro-ph},
       adsurl = {https://ui.adsabs.harvard.edu/abs/2008MNRAS.384.1567M}
}

@ARTICLE{2009ApJ...705.1129L,
       author = {{Lau}, Erwin T. and {Kravtsov}, Andrey V. and {Nagai}, Daisuke},
        title = "{Residual Gas Motions in the Intracluster Medium and Bias in Hydrostatic Measurements of Mass Profiles of Clusters}",
      journal = {\apj},
         year = 2009,
        month = nov,
       volume = {705},
       number = {2},
        pages = {1129-1138},
          doi = {10.1088/0004-637X/705/2/1129},
archivePrefix = {arXiv},
       eprint = {0903.4895},
 primaryClass = {astro-ph.CO},
       adsurl = {https://ui.adsabs.harvard.edu/abs/2009ApJ...705.1129L}
}

@ARTICLE{2007ApJ...655...98N,
       author = {{Nagai}, Daisuke and {Vikhlinin}, Alexey and {Kravtsov}, Andrey V.},
        title = "{Testing X-Ray Measurements of Galaxy Clusters with Cosmological Simulations}",
      journal = {\apj},
         year = 2007,
        month = jan,
       volume = {655},
       number = {1},
        pages = {98-108},
          doi = {10.1086/509868},
archivePrefix = {arXiv},
       eprint = {astro-ph/0609247},
 primaryClass = {astro-ph},
       adsurl = {https://ui.adsabs.harvard.edu/abs/2007ApJ...655...98N}
}

@ARTICLE{2014MNRAS.440.2077M,
       author = {{Mantz}, A.~B. and {Allen}, S.~W. and {Morris}, R.~G. and {Rapetti}, D.~A. and {Applegate}, D.~E. and {Kelly}, P.~L. and {von der Linden}, A. and {Schmidt}, R.~W.},
        title = "{Cosmology and astrophysics from relaxed galaxy clusters - II. Cosmological constraints}",
      journal = {\mnras},
         year = 2014,
        month = may,
       volume = {440},
       number = {3},
        pages = {2077-2098},
          doi = {10.1093/mnras/stu368},
archivePrefix = {arXiv},
       eprint = {1402.6212},
 primaryClass = {astro-ph.CO},
       adsurl = {https://ui.adsabs.harvard.edu/abs/2014MNRAS.440.2077M}
}

@ARTICLE{2009A&A...501...61E,
       author = {{Ettori}, S. and {Morandi}, A. and {Tozzi}, P. and {Balestra}, I. and {Borgani}, S. and {Rosati}, P. and {Lovisari}, L. and {Terenziani}, F.},
        title = "{The cluster gas mass fraction as a cosmological probe: a revised study}",
      journal = {\aap},
         year = 2009,
        month = jul,
       volume = {501},
       number = {1},
        pages = {61-73},
          doi = {10.1051/0004-6361/200810878},
archivePrefix = {arXiv},
       eprint = {0904.2740},
 primaryClass = {astro-ph.CO},
       adsurl = {https://ui.adsabs.harvard.edu/abs/2009A&A...501...61E}
}

@ARTICLE{1974ApJ...187..425P,
       author = {{Press}, William H. and {Schechter}, Paul},
        title = "{Formation of Galaxies and Clusters of Galaxies by Self-Similar Gravitational Condensation}",
      journal = {\apj},
         year = 1974,
        month = feb,
       volume = {187},
        pages = {425-438},
          doi = {10.1086/152650},
       adsurl = {https://ui.adsabs.harvard.edu/abs/1974ApJ...187..425P}
}

@ARTICLE{1993ApJ...407L..49B,
       author = {{Bahcall}, Neta A. and {Cen}, Renyue},
        title = "{The Mass Function of Clusters of Galaxies}",
      journal = {\apjl},
         year = 1993,
        month = apr,
       volume = {407},
        pages = {L49},
          doi = {10.1086/186803},
       adsurl = {https://ui.adsabs.harvard.edu/abs/1993ApJ...407L..49B}
}

@ARTICLE{2001ApJ...553..545H,
       author = {{Haiman}, Zolt{\'a}n and {Mohr}, Joseph J. and {Holder}, Gilbert P.},
        title = "{Constraints on Cosmological Parameters from Future Galaxy Cluster Surveys}",
      journal = {\apj},
         year = 2001,
        month = jun,
       volume = {553},
       number = {2},
        pages = {545-561},
          doi = {10.1086/320939},
archivePrefix = {arXiv},
       eprint = {astro-ph/0002336},
 primaryClass = {astro-ph},
       adsurl = {https://ui.adsabs.harvard.edu/abs/2001ApJ...553..545H}
}

@ARTICLE{1993Natur.366..429W,
       author = {{White}, Simon D.~M. and {Navarro}, Julio F. and {Evrard}, August E. and {Frenk}, Carlos S.},
        title = "{The baryon content of galaxy clusters: a challenge to cosmological orthodoxy}",
      journal = {\nat},
         year = 1993,
        month = dec,
       volume = {366},
       number = {6454},
        pages = {429-433},
          doi = {10.1038/366429a0},
       adsurl = {https://ui.adsabs.harvard.edu/abs/1993Natur.366..429W}
}

@ARTICLE{1996PASJ...48L.119S,
       author = {{Sasaki}, Shin},
        title = "{A New Method to Estimate Cosmological Parameters Using the Baryon Fraction of Clusters of Galaxies}",
      journal = {\pasj},
         year = 1996,
        month = dec,
       volume = {48},
        pages = {L119-L122},
          doi = {10.1093/pasj/48.6.L119},
archivePrefix = {arXiv},
       eprint = {astro-ph/9611033},
 primaryClass = {astro-ph},
       adsurl = {https://ui.adsabs.harvard.edu/abs/1996PASJ...48L.119S}
}

@ARTICLE{1991MNRAS.249..662P,
       author = {{Plionis}, Manolis and {Barrow}, John D. and {Frenk}, Carlos S.},
        title = "{Projected and intrinsic shapes of galaxy clusters.}",
      journal = {\mnras},
         year = 1991,
        month = apr,
       volume = {249},
        pages = {662},
          doi = {10.1093/mnras/249.4.662},
       adsurl = {https://ui.adsabs.harvard.edu/abs/1991MNRAS.249..662P}
}

@ARTICLE{2010ApJ...720.1066C,
       author = {{Cavagnolo}, K.~W. and {McNamara}, B.~R. and {Nulsen}, P.~E.~J. and {Carilli}, C.~L. and {Jones}, C. and {B{\^\i}rzan}, L.},
        title = "{A Relationship Between AGN Jet Power and Radio Power}",
      journal = {\apj},
         year = 2010,
        month = sep,
       volume = {720},
       number = {2},
        pages = {1066-1072},
          doi = {10.1088/0004-637X/720/2/1066},
archivePrefix = {arXiv},
       eprint = {1006.5699},
 primaryClass = {astro-ph.CO},
       adsurl = {https://ui.adsabs.harvard.edu/abs/2010ApJ...720.1066C}
}

@ARTICLE{2022MNRAS.513.3013Y,
       author = {{Yuan}, Z.~S. and {Han}, J.~L. and {Wen}, Z.~L.},
        title = "{Dynamical state of galaxy clusters evaluated from X-ray images}",
      journal = {\mnras},
         year = 2022,
        month = jun,
       volume = {513},
       number = {2},
        pages = {3013-3021},
          doi = {10.1093/mnras/stac1037},
archivePrefix = {arXiv},
       eprint = {2204.02699},
 primaryClass = {astro-ph.GA},
       adsurl = {https://ui.adsabs.harvard.edu/abs/2022MNRAS.513.3013Y}
}

@ARTICLE{2022MNRAS.510..131M,
       author = {{Mantz}, Adam B. and {Morris}, R. Glenn and {Allen}, Steven W. and {Canning}, Rebecca E.~A. and {Baumont}, Lucie and {Benson}, Bradford and {Bleem}, Lindsey E. and {Ehlert}, Steven R. and {Floyd}, Benjamin and {Herbonnet}, Ricardo and {Kelly}, Patrick L. and {Liang}, Shuang and {von der Linden}, Anja and {McDonald}, Michael and {Rapetti}, David A. and {Schmidt}, Robert W. and {Werner}, Norbert and {Wright}, Adam},
        title = "{Cosmological constraints from gas mass fractions of massive, relaxed galaxy clusters}",
      journal = {\mnras},
         year = 2022,
        month = feb,
       volume = {510},
       number = {1},
        pages = {131-145},
          doi = {10.1093/mnras/stab3390},
archivePrefix = {arXiv},
       eprint = {2111.09343},
 primaryClass = {astro-ph.CO},
       adsurl = {https://ui.adsabs.harvard.edu/abs/2022MNRAS.510..131M}
}

@ARTICLE{2023ApJ...947...44C,
       author = {{Calzadilla}, Michael S. and {Bleem}, Lindsey E. and {McDonald}, Michael and {Gladders}, Michael D. and {Mantz}, Adam B. and {Allen}, Steven W. and {Bayliss}, Matthew B. and {Eilers}, Anna-Christina and {Floyd}, Benjamin and {Hlavacek-Larrondo}, Julie and {Khullar}, Gourav and {Kim}, Keunho J. and {Mahler}, Guillaume and {Sharon}, Keren and {Somboonpanyakul}, Taweewat and {Stalder}, Brian and {Stark}, Antony A. and {SPT Collaboration}},
        title = "{SPT-CL J2215-3537: A Massive Starburst at the Center of the Most Distant Relaxed Galaxy Cluster}",
      journal = {\apj},
         year = 2023,
        month = apr,
       volume = {947},
       number = {2},
          eid = {44},
        pages = {44},
          doi = {10.3847/1538-4357/acc6c2},
archivePrefix = {arXiv},
       eprint = {2303.10185},
 primaryClass = {astro-ph.GA},
       adsurl = {https://ui.adsabs.harvard.edu/abs/2023ApJ...947...44C}
}

@ARTICLE{1996ApJ...458...27B,
       author = {{Buote}, David A. and {Tsai}, John C.},
        title = "{Quantifying the Morphologies and Dynamical Evolution of Galaxy Clusters. II. Application to a Sample of ROSAT Clusters}",
      journal = {\apj},
         year = 1996,
        month = feb,
       volume = {458},
        pages = {27},
          doi = {10.1086/176790},
archivePrefix = {arXiv},
       eprint = {astro-ph/9504046},
 primaryClass = {astro-ph},
       adsurl = {https://ui.adsabs.harvard.edu/abs/1996ApJ...458...27B}
}

@article{10.1093/mnras/staa2363,
    author = {{Yuan}, Z.~S. and {Han}, J.~L.},
    title = {Dynamical state for 964 galaxy clusters from Chandra X-ray images},
    journal = {Monthly Notices of the Royal Astronomical Society},
    volume = {497},
    number = {4},
    pages = {5485-5497},
    year = {2020},
    month = {08},
    issn = {0035-8711},
    doi = {10.1093/mnras/staa2363},
    url = {https://doi.org/10.1093/mnras/staa2363},
    eprint = {https://academic.oup.com/mnras/article-pdf/497/4/5485/33695831/staa2363.pdf},
}

@ARTICLE{2006MNRAS.373..881P,
       author = {{Poole}, Gregory B. and {Fardal}, Mark A. and {Babul}, Arif and {McCarthy}, Ian G. and {Quinn}, Thomas and {Wadsley}, James},
        title = "{The impact of mergers on relaxed X-ray clusters - I. Dynamical evolution and emergent transient structures}",
      journal = {\mnras},
         year = 2006,
        month = dec,
       volume = {373},
       number = {3},
        pages = {881-905},
          doi = {10.1111/j.1365-2966.2006.10916.x},
archivePrefix = {arXiv},
       eprint = {astro-ph/0608560},
 primaryClass = {astro-ph},
       adsurl = {https://ui.adsabs.harvard.edu/abs/2006MNRAS.373..881P}
}

@ARTICLE{1995ApJ...452..522B,
       author = {{Buote}, David A. and {Tsai}, John C.},
        title = "{Quantifying the Morphologies and Dynamical Evolution of Galaxy Clusters. I. The Method}",
      journal = {\apj},
         year = 1995,
        month = oct,
       volume = {452},
        pages = {522},
          doi = {10.1086/176326},
archivePrefix = {arXiv},
       eprint = {astro-ph/9502002},
 primaryClass = {astro-ph},
       adsurl = {https://ui.adsabs.harvard.edu/abs/1995ApJ...452..522B}
}

@ARTICLE{2006ApJ...650..102A,
       author = {{Ascasibar}, Yago and {Markevitch}, Maxim},
        title = "{The Origin of Cold Fronts in the Cores of Relaxed Galaxy Clusters}",
      journal = {\apj},
         year = 2006,
        month = oct,
       volume = {650},
       number = {1},
        pages = {102-127},
          doi = {10.1086/506508},
archivePrefix = {arXiv},
       eprint = {astro-ph/0603246},
 primaryClass = {astro-ph},
       adsurl = {https://ui.adsabs.harvard.edu/abs/2006ApJ...650..102A}
}

@ARTICLE{2010A&A...516A..32G,
       author = {{Ghizzardi}, S. and {Rossetti}, M. and {Molendi}, S.},
        title = "{Cold fronts in galaxy clusters}",
      journal = {\aap},
         year = 2010,
        month = jun,
       volume = {516},
          eid = {A32},
        pages = {A32},
          doi = {10.1051/0004-6361/200912496},
archivePrefix = {arXiv},
       eprint = {1003.1051},
 primaryClass = {astro-ph.CO},
       adsurl = {https://ui.adsabs.harvard.edu/abs/2010A&A...516A..32G}
}

@ARTICLE{2016ApJS..227...31S,
       author = {{Shin}, Jaejin and {Woo}, Jong-Hak and {Mulchaey}, John S.},
        title = "{A Systematic Search for X-Ray Cavities in Galaxy Clusters, Groups, and Elliptical Galaxies}",
      journal = {\apjs},
         year = 2016,
        month = dec,
       volume = {227},
       number = {2},
          eid = {31},
        pages = {31},
          doi = {10.3847/1538-4365/227/2/31},
archivePrefix = {arXiv},
       eprint = {1610.03487},
 primaryClass = {astro-ph.CO},
       adsurl = {https://ui.adsabs.harvard.edu/abs/2016ApJS..227...31S}
}

@ARTICLE{2023Galax..11...73B,
       author = {{Bourne}, Martin A. and {Yang}, Hsiang-Yi Karen},
        title = "{Recent Progress in Modeling the Macro- and Micro-Physics of Radio Jet Feedback in Galaxy Clusters}",
      journal = {Galaxies},
         year = 2023,
        month = jun,
       volume = {11},
       number = {3},
          eid = {73},
        pages = {73},
          doi = {10.3390/galaxies11030073},
archivePrefix = {arXiv},
       eprint = {2305.00019},
 primaryClass = {astro-ph.HE},
       adsurl = {https://ui.adsabs.harvard.edu/abs/2023Galax..11...73B}
}

@ARTICLE{2021MNRAS.503.3394C,
       author = {{Cao}, Kaili and {Barnes}, David J. and {Vogelsberger}, Mark},
        title = "{Studying galaxy cluster morphological metrics with MOCK-X}",
      journal = {\mnras},
         year = 2021,
        month = may,
       volume = {503},
       number = {3},
        pages = {3394-3413},
          doi = {10.1093/mnras/stab605},
archivePrefix = {arXiv},
       eprint = {2006.10752},
 primaryClass = {astro-ph.CO},
       adsurl = {https://ui.adsabs.harvard.edu/abs/2021MNRAS.503.3394C}
}

@ARTICLE{2008ApJS..174..117M,
       author = {{Maughan}, B.~J. and {Jones}, C. and {Forman}, W. and {Van Speybroeck}, L.},
        title = "{Images, Structural Properties, and Metal Abundances of Galaxy Clusters Observed with Chandra ACIS-I at 0.1 < z < 1.3}",
      journal = {\apjs},
         year = 2008,
        month = jan,
       volume = {174},
       number = {1},
        pages = {117-135},
          doi = {10.1086/521225},
archivePrefix = {arXiv},
       eprint = {astro-ph/0703156},
 primaryClass = {astro-ph},
       adsurl = {https://ui.adsabs.harvard.edu/abs/2008ApJS..174..117M}
}

@ARTICLE{2024ApJ...976..169C,
       author = {{Calzadilla}, Michael S. and {McDonald}, Michael and {Benson}, Bradford A. and {Bleem}, Lindsey E. and {Croston}, Judith H. and {Donahue}, Megan and {Edge}, Alastair C. and {Floyd}, Benjamin and {Garmire}, Gordon P. and {Hlavacek-Larrondo}, Julie and {Huynh}, Minh T. and {Khullar}, Gourav and {Kraft}, Ralph P. and {McNamara}, Brian R. and {Noble}, Allison G. and {Romero}, Charles E. and {Ruppin}, Florian and {Somboonpanyakul}, Taweewat and {Voit}, G. Mark},
        title = "{The SPT-Chandra BCG Spectroscopic Survey. I. Evolution of the Entropy Threshold for ICM Cooling and AGN Feedback in Galaxy Clusters over the Last 10 Gyr}",
      journal = {\apj},
         year = 2024,
        month = dec,
       volume = {976},
       number = {2},
          eid = {169},
        pages = {169},
          doi = {10.3847/1538-4357/ad8916},
archivePrefix = {arXiv},
       eprint = {2311.00396},
 primaryClass = {astro-ph.GA},
       adsurl = {https://ui.adsabs.harvard.edu/abs/2024ApJ...976..169C}
}

@ARTICLE{2002MNRAS.334L..11A,
       author = {{Allen}, S.~W. and {Schmidt}, R.~W. and {Fabian}, A.~C.},
        title = "{Cosmological constraints from the X-ray gas mass fraction in relaxed lensing clusters observed with Chandra}",
      journal = {\mnras},
         year = 2002,
        month = aug,
       volume = {334},
       number = {2},
        pages = {L11-L15},
          doi = {10.1046/j.1365-8711.2002.05601.x},
archivePrefix = {arXiv},
       eprint = {astro-ph/0205007},
 primaryClass = {astro-ph},
       adsurl = {https://ui.adsabs.harvard.edu/abs/2002MNRAS.334L..11A}
}

@ARTICLE{2016A&A...594A.116H,
       author = {{HI4PI Collaboration} and {Ben Bekhti}, N. and {Fl{\"o}er}, L. and {Keller}, R. and {Kerp}, J. and {Lenz}, D. and {Winkel}, B. and {Bailin}, J. and {Calabretta}, M.~R. and {Dedes}, L. and {Ford}, H.~A. and {Gibson}, B.~K. and {Haud}, U. and {Janowiecki}, S. and {Kalberla}, P.~M.~W. and {Lockman}, F.~J. and {McClure-Griffiths}, N.~M. and {Murphy}, T. and {Nakanishi}, H. and {Pisano}, D.~J. and {Staveley-Smith}, L.},
        title = "{HI4PI: A full-sky H I survey based on EBHIS and GASS}",
      journal = {\aap},
         year = 2016,
        month = oct,
       volume = {594},
          eid = {A116},
        pages = {A116},
          doi = {10.1051/0004-6361/201629178},
archivePrefix = {arXiv},
       eprint = {1610.06175},
 primaryClass = {astro-ph.GA},
       adsurl = {https://ui.adsabs.harvard.edu/abs/2016A&A...594A.116H}
}

@ARTICLE{2005ApJ...624..606J,
       author = {{Jeltema}, Tesla E. and {Canizares}, Claude R. and {Bautz}, Mark W. and {Buote}, David A.},
        title = "{The Evolution of Structure in X-Ray Clusters of Galaxies}",
      journal = {\apj},
         year = 2005,
        month = may,
       volume = {624},
       number = {2},
        pages = {606-629},
          doi = {10.1086/428940},
archivePrefix = {arXiv},
       eprint = {astro-ph/0501360},
 primaryClass = {astro-ph},
       adsurl = {https://ui.adsabs.harvard.edu/abs/2005ApJ...624..606J}
}

@ARTICLE{2001ApJ...546...63T,
       author = {{Tozzi}, Paolo and {Norman}, Colin},
        title = "{The Evolution of X-Ray Clusters and the Entropy of the Intracluster Medium}",
      journal = {\apj},
         year = 2001,
        month = jan,
       volume = {546},
       number = {1},
        pages = {63-84},
          doi = {10.1086/318237},
archivePrefix = {arXiv},
       eprint = {astro-ph/0003289},
 primaryClass = {astro-ph},
       adsurl = {https://ui.adsabs.harvard.edu/abs/2001ApJ...546...63T}
}

@ARTICLE{2005MNRAS.364..909V,
       author = {{Voit}, G. Mark and {Kay}, Scott T. and {Bryan}, Greg L.},
        title = "{The baseline intracluster entropy profile from gravitational structure formation}",
      journal = {\mnras},
         year = 2005,
        month = dec,
       volume = {364},
       number = {3},
        pages = {909-916},
          doi = {10.1111/j.1365-2966.2005.09621.x},
archivePrefix = {arXiv},
       eprint = {astro-ph/0511252},
 primaryClass = {astro-ph},
       adsurl = {https://ui.adsabs.harvard.edu/abs/2005MNRAS.364..909V}
}

@ARTICLE{2009ApJS..182...12C,
       author = {{Cavagnolo}, Kenneth W. and {Donahue}, Megan and {Voit}, G. Mark and {Sun}, Ming},
        title = "{Intracluster Medium Entropy Profiles for a Chandra Archival Sample of Galaxy Clusters}",
      journal = {\apjs},
         year = 2009,
        month = may,
       volume = {182},
       number = {1},
        pages = {12-32},
          doi = {10.1088/0067-0049/182/1/12},
archivePrefix = {arXiv},
       eprint = {0902.1802},
 primaryClass = {astro-ph.CO},
       adsurl = {https://ui.adsabs.harvard.edu/abs/2009ApJS..182...12C}
}

@ARTICLE{2021MNRAS.506.2533B,
       author = {{Barnes}, David J. and {Vogelsberger}, Mark and {Pearce}, Francesca A. and {Pop}, Ana-Roxana and {Kannan}, Rahul and {Cao}, Kaili and {Kay}, Scott T. and {Hernquist}, Lars},
        title = "{Characterizing hydrostatic mass bias with MOCK-X}",
      journal = {\mnras},
         year = 2021,
        month = sep,
       volume = {506},
       number = {2},
        pages = {2533-2550},
          doi = {10.1093/mnras/stab1276},
archivePrefix = {arXiv},
       eprint = {2001.11508},
 primaryClass = {astro-ph.CO},
       adsurl = {https://ui.adsabs.harvard.edu/abs/2021MNRAS.506.2533B}
}

@ARTICLE{2023ApJ...945...71L,
       author = {{Lee}, Wonki and {Cha}, Sangjun and {Jee}, M. James and {Nagai}, Daisuke and {King}, Lindsay and {ZuHone}, John and {Chadayammuri}, Urmila and {Felix}, Sharon and {Finner}, Kyle},
        title = "{Weak-lensing Mass Bias in Merging Galaxy Clusters}",
      journal = {\apj},
         year = 2023,
        month = mar,
       volume = {945},
       number = {1},
          eid = {71},
        pages = {71},
          doi = {10.3847/1538-4357/acb76b},
archivePrefix = {arXiv},
       eprint = {2211.03892},
 primaryClass = {astro-ph.CO},
       adsurl = {https://ui.adsabs.harvard.edu/abs/2023ApJ...945...71L}
}

@INPROCEEDINGS{1996ASPC..101...17A,
       author = {{Arnaud}, K.~A.},
        title = "{XSPEC: The First Ten Years}",
    booktitle = {Astronomical Data Analysis Software and Systems V},
         year = 1996,
       editor = {{Jacoby}, George H. and {Barnes}, Jeannette},
       series = {Astronomical Society of the Pacific Conference Series},
       volume = {101},
        month = jan,
        pages = {17},
       adsurl = {https://ui.adsabs.harvard.edu/abs/1996ASPC..101...17A}
}

@ARTICLE{2015Natur.519..203V,
       author = {{Voit}, G.~M. and {Donahue}, M. and {Bryan}, G.~L. and {McDonald}, M.},
        title = "{Regulation of star formation in giant galaxies by precipitation, feedback and conduction}",
      journal = {\nat},
         year = 2015,
        month = mar,
       volume = {519},
       number = {7542},
        pages = {203-206},
          doi = {10.1038/nature14167},
archivePrefix = {arXiv},
       eprint = {1409.1598},
 primaryClass = {astro-ph.GA},
       adsurl = {https://ui.adsabs.harvard.edu/abs/2015Natur.519..203V}
}

@ARTICLE{2017ApJ...845...80V,
       author = {{Voit}, G. Mark and {Meece}, Greg and {Li}, Yuan and {O'Shea}, Brian W. and {Bryan}, Greg L. and {Donahue}, Megan},
        title = "{A Global Model for Circumgalactic and Cluster-core Precipitation}",
      journal = {\apj},
         year = 2017,
        month = aug,
       volume = {845},
       number = {1},
          eid = {80},
        pages = {80},
          doi = {10.3847/1538-4357/aa7d04},
archivePrefix = {arXiv},
       eprint = {1607.02212},
 primaryClass = {astro-ph.GA},
       adsurl = {https://ui.adsabs.harvard.edu/abs/2017ApJ...845...80V}
}

@ARTICLE{2013ApJ...762...69Z,
       author = {{ZuHone}, J.~A. and {Markevitch}, M. and {Ruszkowski}, M. and {Lee}, D.},
        title = "{Cold Fronts and Gas Sloshing in Galaxy Clusters with Anisotropic Thermal Conduction}",
      journal = {\apj},
         year = 2013,
        month = jan,
       volume = {762},
       number = {2},
          eid = {69},
        pages = {69},
          doi = {10.1088/0004-637X/762/2/69},
archivePrefix = {arXiv},
       eprint = {1204.6005},
 primaryClass = {astro-ph.CO},
       adsurl = {https://ui.adsabs.harvard.edu/abs/2013ApJ...762...69Z}
}

@ARTICLE{2011ApJ...743...16Z,
       author = {{ZuHone}, J.~A. and {Markevitch}, M. and {Lee}, D.},
        title = "{Sloshing of the Magnetized Cool Gas in the Cores of Galaxy Clusters}",
      journal = {\apj},
         year = 2011,
        month = dec,
       volume = {743},
       number = {1},
          eid = {16},
        pages = {16},
          doi = {10.1088/0004-637X/743/1/16},
archivePrefix = {arXiv},
       eprint = {1108.4427},
 primaryClass = {astro-ph.CO},
       adsurl = {https://ui.adsabs.harvard.edu/abs/2011ApJ...743...16Z}
}

@ARTICLE{2017MNRAS.464.4360M,
       author = {{Main}, R.~A. and {McNamara}, B.~R. and {Nulsen}, P.~E.~J. and {Russell}, H.~R. and {Vantyghem}, A.~N.},
        title = "{A relationship between halo mass, cooling, active galactic nuclei heating and the co-evolution of massive black holes}",
      journal = {\mnras},
         year = 2017,
        month = feb,
       volume = {464},
       number = {4},
        pages = {4360-4382},
          doi = {10.1093/mnras/stw2644},
archivePrefix = {arXiv},
       eprint = {1510.07046},
 primaryClass = {astro-ph.GA},
       adsurl = {https://ui.adsabs.harvard.edu/abs/2017MNRAS.464.4360M}
}

@ARTICLE{2006MNRAS.373..959D,
       author = {{Dunn}, R.~J.~H. and {Fabian}, A.~C.},
        title = "{Investigating AGN heating in a sample of nearby clusters}",
      journal = {\mnras},
         year = 2006,
        month = dec,
       volume = {373},
       number = {3},
        pages = {959-971},
          doi = {10.1111/j.1365-2966.2006.11080.x},
archivePrefix = {arXiv},
       eprint = {astro-ph/0609537},
 primaryClass = {astro-ph},
       adsurl = {https://ui.adsabs.harvard.edu/abs/2006MNRAS.373..959D}
}

@ARTICLE{2015ApJ...811...73L,
       author = {{Li}, Yuan and {Bryan}, Greg L. and {Ruszkowski}, Mateusz and {Voit}, G. Mark and {O'Shea}, Brian W. and {Donahue}, Megan},
        title = "{Cooling, AGN Feedback, and Star Formation in Simulated Cool-core Galaxy Clusters}",
      journal = {\apj},
         year = 2015,
        month = oct,
       volume = {811},
       number = {2},
          eid = {73},
        pages = {73},
          doi = {10.1088/0004-637X/811/2/73},
archivePrefix = {arXiv},
       eprint = {1503.02660},
 primaryClass = {astro-ph.GA},
       adsurl = {https://ui.adsabs.harvard.edu/abs/2015ApJ...811...73L}
}

@ARTICLE{2024Univ...10..273R,
       author = {{Russell}, Helen R. and {Lopez}, Laura A. and {Allen}, Steven W. and {Chartas}, George and {Choudhury}, Prakriti Pal and {Dupke}, Renato A. and {Fabian}, Andrew C. and {Flores}, Anthony M. and {Garofali}, Kristen and {Hodges-Kluck}, Edmund and {Koss}, Michael J. and {Lanz}, Lauranne and {Lehmer}, Bret D. and {Li}, Jiang-Tao and {Maksym}, W. Peter and {Mantz}, Adam B. and {McDonald}, Michael and {Miller}, Eric D. and {Mushotzky}, Richard F. and {Qiu}, Yu and {Reynolds}, Christopher S. and {Tombesi}, Francesco and {Tozzi}, Paolo and {Trindade-Falc{\~a}o}, Anna and {Walker}, Stephen A. and {Wong}, Ka-Wah and {Yukita}, Mihoko and {Zhang}, Congyao},
        title = "{The Evolution of Galaxies and Clusters at High Spatial Resolution with Advanced X-ray Imaging Satellite (AXIS)}",
      journal = {Universe},
         year = 2024,
        month = jun,
       volume = {10},
       number = {7},
          eid = {273},
        pages = {273},
          doi = {10.3390/universe10070273},
archivePrefix = {arXiv},
       eprint = {2311.07661},
 primaryClass = {astro-ph.IM},
       adsurl = {https://ui.adsabs.harvard.edu/abs/2024Univ...10..273R}
}

@ARTICLE{2004ApJ...616..178C,
       author = {{Clarke}, T.~E. and {Blanton}, Elizabeth L. and {Sarazin}, Craig L.},
        title = "{The Complex Cooling Core of A2029: Radio and X-Ray Interactions}",
      journal = {\apj},
         year = 2004,
        month = nov,
       volume = {616},
       number = {1},
        pages = {178-191},
          doi = {10.1086/424911},
archivePrefix = {arXiv},
       eprint = {astro-ph/0408068},
 primaryClass = {astro-ph},
       adsurl = {https://ui.adsabs.harvard.edu/abs/2004ApJ...616..178C}
}

@ARTICLE{2005MNRAS.360L..20F,
       author = {{Fabian}, A.~C. and {Sanders}, J.~S. and {Taylor}, G.~B. and {Allen}, S.~W.},
        title = "{A deep Chandra observation of the Centaurus cluster: bubbles, filaments and edges}",
      journal = {\mnras},
         year = 2005,
        month = jun,
       volume = {360},
       number = {1},
        pages = {L20-L24},
          doi = {10.1111/j.1745-3933.2005.00037.x},
archivePrefix = {arXiv},
       eprint = {astro-ph/0503154},
 primaryClass = {astro-ph},
       adsurl = {https://ui.adsabs.harvard.edu/abs/2005MNRAS.360L..20F}
}

@ARTICLE{2007ApJ...671..181D,
       author = {{Dupke}, Renato and {White}, III, Raymond E. and {Bregman}, Joel N.},
        title = "{Different Methods of Forming Cold Fronts in Nonmerging Clusters}",
      journal = {\apj},
         year = 2007,
        month = dec,
       volume = {671},
       number = {1},
        pages = {181-189},
          doi = {10.1086/522194},
archivePrefix = {arXiv},
       eprint = {0707.4001},
 primaryClass = {astro-ph},
       adsurl = {https://ui.adsabs.harvard.edu/abs/2007ApJ...671..181D}
}

@ARTICLE{2004ApJ...609..638G,
       author = {{Ghizzardi}, Simona and {Molendi}, Silvano and {Pizzolato}, Fabio and {De Grandi}, Sabrina},
        title = "{Radiative Cooling and Heating and Thermal Conduction in M87}",
      journal = {\apj},
         year = 2004,
        month = jul,
       volume = {609},
       number = {2},
        pages = {638-651},
          doi = {10.1086/421314},
archivePrefix = {arXiv},
       eprint = {astro-ph/0404060},
 primaryClass = {astro-ph},
       adsurl = {https://ui.adsabs.harvard.edu/abs/2004ApJ...609..638G}
}

@ARTICLE{2016A&ARv..24...14O,
       author = {{Overzier}, Roderik A.},
        title = "{The realm of the galaxy protoclusters. A review}",
      journal = {\aapr},
         year = 2016,
        month = nov,
       volume = {24},
       number = {1},
          eid = {14},
        pages = {14},
          doi = {10.1007/s00159-016-0100-3},
archivePrefix = {arXiv},
       eprint = {1610.05201},
 primaryClass = {astro-ph.GA},
       adsurl = {https://ui.adsabs.harvard.edu/abs/2016A&ARv..24...14O}
}

@ARTICLE{1989GeCoA..53..197A,
       author = {{Anders}, E. and {Grevesse}, N.},
        title = "{Abundances of the elements: Meteoritic and solar}",
      journal = {\gca},
         year = 1989,
        month = jan,
       volume = {53},
       number = {1},
        pages = {197-214},
          doi = {10.1016/0016-7037(89)90286-X},
       adsurl = {https://ui.adsabs.harvard.edu/abs/1989GeCoA..53..197A}
}

@ARTICLE{2001MNRAS.325..178S,
       author = {{Sanders}, J.~S. and {Fabian}, A.~C.},
        title = "{Adaptive binning of X-ray galaxy cluster images}",
      journal = {\mnras},
         year = 2001,
        month = jul,
       volume = {325},
       number = {1},
        pages = {178-186},
          doi = {10.1046/j.1365-8711.2001.04410.x},
archivePrefix = {arXiv},
       eprint = {astro-ph/0011500},
 primaryClass = {astro-ph},
       adsurl = {https://ui.adsabs.harvard.edu/abs/2001MNRAS.325..178S}
}

@ARTICLE{2024ApJ...960L...7J,
       author = {{Jim{\'e}nez-Teja}, Yolanda and {Dupke}, Renato A. and {Lopes}, Paulo A.~A. and {Dimauro}, Paola},
        title = "{Evidence for a Redshifted Excess in the Intracluster Light Fractions of Merging Clusters at z   0.8}",
      journal = {\apjl},
         year = 2024,
        month = jan,
       volume = {960},
       number = {2},
          eid = {L7},
        pages = {L7},
          doi = {10.3847/2041-8213/ad181a},
archivePrefix = {arXiv},
       eprint = {2401.02543},
 primaryClass = {astro-ph.GA},
       adsurl = {https://ui.adsabs.harvard.edu/abs/2024ApJ...960L...7J}
}

@ARTICLE{ned,
       author = {{Wright}, E.~L.},
        title = "{A Cosmology Calculator for the World Wide Web}",
      journal = {\pasp},
         year = 2006,
        month = dec,
       volume = {118},
       number = {850},
        pages = {1711-1715},
          doi = {10.1086/510102},
archivePrefix = {arXiv},
       eprint = {astro-ph/0609593},
 primaryClass = {astro-ph},
       adsurl = {https://ui.adsabs.harvard.edu/abs/2006PASP..118.1711W}
}

@ARTICLE{2023A&A...675A.188A,
       author = {{Angelinelli}, M. and {Ettori}, S. and {Dolag}, K. and {Vazza}, F. and {Ragagnin}, A.},
        title = "{Redshift evolution of the baryon and gas fraction in simulated groups and clusters of galaxies}",
      journal = {\aap},
         year = 2023,
        month = jul,
       volume = {675},
          eid = {A188},
        pages = {A188},
          doi = {10.1051/0004-6361/202245782},
archivePrefix = {arXiv},
       eprint = {2305.09733},
 primaryClass = {astro-ph.CO},
       adsurl = {https://ui.adsabs.harvard.edu/abs/2023A&A...675A.188A}
}

@ARTICLE{2024arXiv241116555P,
       author = {{Popesso}, P. and {Biviano}, A. and {Marini}, I. and {Dolag}, K. and {Vladutescu-Zopp}, S. and {Csizi}, B. and {Biffi}, V. and {Lamer}, G. and {Robothan}, A. and {Bravo}, M. and {Lovisari}, L. and {Ettori}, S. and {Angelinelli}, M. and {Driver}, S. and {Toptun}, V. and {Dev}, A. and {Mazengo}, D. and {Merloni}, A. and {Comparat}, J. and {Ponti}, G. and {Mroczkowski}, T. and {Bulbul}, E. and {Grandis}, S. and {Bahar}, E.},
        title = "{The hot gas mass fraction in halos. From Milky Way-like groups to massive clusters}",
      journal = {arXiv e-prints},
         year = 2024,
        month = nov,
          eid = {arXiv:2411.16555},
        pages = {arXiv:2411.16555},
          doi = {10.48550/arXiv.2411.16555},
archivePrefix = {arXiv},
       eprint = {2411.16555},
 primaryClass = {astro-ph.GA},
       adsurl = {https://ui.adsabs.harvard.edu/abs/2024arXiv241116555P}
}

@ARTICLE{2006ApJ...640..691V,
       author = {{Vikhlinin}, A. and {Kravtsov}, A. and {Forman}, W. and {Jones}, C. and {Markevitch}, M. and {Murray}, S.~S. and {Van Speybroeck}, L.},
        title = "{Chandra Sample of Nearby Relaxed Galaxy Clusters: Mass, Gas Fraction, and Mass-Temperature Relation}",
      journal = {\apj},
         year = 2006,
        month = apr,
       volume = {640},
       number = {2},
        pages = {691-709},
          doi = {10.1086/500288},
archivePrefix = {arXiv},
       eprint = {astro-ph/0507092},
 primaryClass = {astro-ph},
       adsurl = {https://ui.adsabs.harvard.edu/abs/2006ApJ...640..691V}
}

@ARTICLE{2016MNRAS.461.4182M,
       author = {{Maughan}, Ben J. and {Giles}, Paul A. and {Rines}, Kenneth J. and {Diaferio}, Antonaldo and {Geller}, Margaret J. and {Van Der Pyl}, Nina and {Bonamente}, Massimiliano},
        title = "{Hydrostatic and caustic mass profiles of galaxy clusters}",
      journal = {\mnras},
         year = 2016,
        month = oct,
       volume = {461},
       number = {4},
        pages = {4182-4191},
          doi = {10.1093/mnras/stw1610},
archivePrefix = {arXiv},
       eprint = {1511.07872},
 primaryClass = {astro-ph.CO},
       adsurl = {https://ui.adsabs.harvard.edu/abs/2016MNRAS.461.4182M}
}

@ARTICLE{1997ApJ...481..633D,
       author = {{Diaferio}, Antonaldo and {Geller}, Margaret J.},
        title = "{Infall Regions of Galaxy Clusters}",
      journal = {\apj},
         year = 1997,
        month = may,
       volume = {481},
       number = {2},
        pages = {633-643},
          doi = {10.1086/304075},
archivePrefix = {arXiv},
       eprint = {astro-ph/9701034},
 primaryClass = {astro-ph},
       adsurl = {https://ui.adsabs.harvard.edu/abs/1997ApJ...481..633D}
}

@ARTICLE{1999MNRAS.309..610D,
       author = {{Diaferio}, Antonaldo},
        title = "{Mass estimation in the outer regions of galaxy clusters}",
      journal = {\mnras},
         year = 1999,
        month = nov,
       volume = {309},
       number = {3},
        pages = {610-622},
          doi = {10.1046/j.1365-8711.1999.02864.x},
archivePrefix = {arXiv},
       eprint = {astro-ph/9906331},
 primaryClass = {astro-ph},
       adsurl = {https://ui.adsabs.harvard.edu/abs/1999MNRAS.309..610D}
}

@ARTICLE{2009arXiv0901.0868D,
       author = {{Diaferio}, Antonaldo},
        title = "{Measuring the mass profile of galaxy clusters beyond their virial radius}",
      journal = {arXiv e-prints},
         year = 2009,
        month = jan,
          eid = {arXiv:0901.0868},
        pages = {arXiv:0901.0868},
          doi = {10.48550/arXiv.0901.0868},
archivePrefix = {arXiv},
       eprint = {0901.0868},
 primaryClass = {astro-ph.CO},
       adsurl = {https://ui.adsabs.harvard.edu/abs/2009arXiv0901.0868D}
}

@ARTICLE{2016MNRAS.457.1522A,
       author = {{Applegate}, D.~E. and {Mantz}, A. and {Allen}, S.~W. and {von der Linden}, A. and {Morris}, R. Glenn and {Hilbert}, S. and {Kelly}, Patrick L. and {Burke}, D.~L. and {Ebeling}, H. and {Rapetti}, D.~A. and {Schmidt}, R.~W.},
        title = "{Cosmology and astrophysics from relaxed galaxy clusters - IV. Robustly calibrating hydrostatic masses with weak lensing}",
      journal = {\mnras},
         year = 2016,
        month = apr,
       volume = {457},
       number = {2},
        pages = {1522-1534},
          doi = {10.1093/mnras/stw005},
archivePrefix = {arXiv},
       eprint = {1509.02162},
 primaryClass = {astro-ph.CO},
       adsurl = {https://ui.adsabs.harvard.edu/abs/2016MNRAS.457.1522A}
}

@misc{pykrige,
  doi = {10.5281/ZENODO.11360184},
  url = {https://zenodo.org/doi/10.5281/zenodo.11360184},
  author = {Benjamin Murphy and Roman Yurchak and Sebastian M\"{u}ller},
  language = {en},
  title = {GeoStat-Framework/PyKrige: v1.7.2},
  publisher = {Zenodo},
  year = {2024},
  copyright = {BSD 3-Clause "New" or "Revised" License}
}

@INPROCEEDINGS{2006SPIE.6270E..1VF,
       author = {{Fruscione}, Antonella and {McDowell}, Jonathan C. and {Allen}, Glenn E. and {Brickhouse}, Nancy S. and {Burke}, Douglas J. and {Davis}, John E. and {Durham}, Nick and {Elvis}, Martin and {Galle}, Elizabeth C. and {Harris}, Daniel E. and {Huenemoerder}, David P. and {Houck}, John C. and {Ishibashi}, Bish and {Karovska}, Margarita and {Nicastro}, Fabrizio and {Noble}, Michael S. and {Nowak}, Michael A. and {Primini}, Frank A. and {Siemiginowska}, Aneta and {Smith}, Randall K. and {Wise}, Michael},
        title = "{CIAO: Chandra's data analysis system}",
    booktitle = {Observatory Operations: Strategies, Processes, and Systems},
         year = 2006,
       editor = {{Silva}, David R. and {Doxsey}, Rodger E.},
       series = {Society of Photo-Optical Instrumentation Engineers (SPIE) Conference Series},
       volume = {6270},
        month = jun,
          eid = {62701V},
        pages = {62701V},
          doi = {10.1117/12.671760},
       adsurl = {https://ui.adsabs.harvard.edu/abs/2006SPIE.6270E..1VF}
}

@ARTICLE{2020ApJS..247...25B,
       author = {{Bleem}, L.~E. and {Bocquet}, S. and {Stalder}, B. and {Gladders}, M.~D. and {Ade}, P.~A.~R. and {Allen}, S.~W. and {Anderson}, A.~J. and {Annis}, J. and {Ashby}, M.~L.~N. and {Austermann}, J.~E. and {Avila}, S. and {Avva}, J.~S. and {Bayliss}, M. and {Beall}, J.~A. and {Bechtol}, K. and {Bender}, A.~N. and {Benson}, B.~A. and {Bertin}, E. and {Bianchini}, F. and {Blake}, C. and {Brodwin}, M. and {Brooks}, D. and {Buckley-Geer}, E. and {Burke}, D.~L. and {Carlstrom}, J.~E. and {Rosell}, A. Carnero and {Carrasco Kind}, M. and {Carretero}, J. and {Chang}, C.~L. and {Chiang}, H.~C. and {Citron}, R. and {Moran}, C. Corbett and {Costanzi}, M. and {Crawford}, T.~M. and {Crites}, A.~T. and {da Costa}, L.~N. and {de Haan}, T. and {De Vicente}, J. and {Desai}, S. and {Diehl}, H.~T. and {Dietrich}, J.~P. and {Dobbs}, M.~A. and {Eifler}, T.~F. and {Everett}, W. and {Flaugher}, B. and {Floyd}, B. and {Frieman}, J. and {Gallicchio}, J. and {Garc{\'\i}a-Bellido}, J. and {George}, E.~M. and {Gerdes}, D.~W. and {Gilbert}, A. and {Gruen}, D. and {Gruendl}, R.~A. and {Gschwend}, J. and {Gupta}, N. and {Gutierrez}, G. and {Halverson}, N.~W. and {Harrington}, N. and {Henning}, J.~W. and {Heymans}, C. and {Holder}, G.~P. and {Hollowood}, D.~L. and {Holzapfel}, W.~L. and {Honscheid}, K. and {Hrubes}, J.~D. and {Huang}, N. and {Hubmayr}, J. and {Irwin}, K.~D. and {James}, D.~J. and {Jeltema}, T. and {Joudaki}, S. and {Khullar}, G. and {Klein}, M. and {Knox}, L. and {Kuropatkin}, N. and {Lee}, A.~T. and {Li}, D. and {Lidman}, C. and {Lowitz}, A. and {MacCrann}, N. and {Mahler}, G. and {Maia}, M.~A.~G. and {Marshall}, J.~L. and {McDonald}, M. and {McMahon}, J.~J. and {Melchior}, P. and {Menanteau}, F. and {Meyer}, S.~S. and {Miquel}, R. and {Mocanu}, L.~M. and {Mohr}, J.~J. and {Montgomery}, J. and {Nadolski}, A. and {Natoli}, T. and {Nibarger}, J.~P. and {Noble}, G. and {Novosad}, V. and {Padin}, S. and {Palmese}, A. and {Parkinson}, D. and {Patil}, S. and {Paz-Chinch{\'o}n}, F. and {Plazas}, A.~A. and {Pryke}, C. and {Ramachandra}, N.~S. and {Reichardt}, C.~L. and {Remolina Gonz{\'a}lez}, J.~D. and {Romer}, A.~K. and {Roodman}, A. and {Ruhl}, J.~E. and {Rykoff}, E.~S. and {Saliwanchik}, B.~R. and {Sanchez}, E. and {Saro}, A. and {Sayre}, J.~T. and {Schaffer}, K.~K. and {Schrabback}, T. and {Serrano}, S. and {Sharon}, K. and {Sievers}, C. and {Smecher}, G. and {Smith}, M. and {Soares-Santos}, M. and {Stark}, A.~A. and {Story}, K.~T. and {Suchyta}, E. and {Tarle}, G. and {Tucker}, C. and {Vanderlinde}, K. and {Veach}, T. and {Vieira}, J.~D. and {Wang}, G. and {Weller}, J. and {Whitehorn}, N. and {Wu}, W.~L.~K. and {Yefremenko}, V. and {Zhang}, Y.},
        title = "{The SPTpol Extended Cluster Survey}",
      journal = {\apjs},
         year = 2020,
        month = mar,
       volume = {247},
       number = {1},
          eid = {25},
        pages = {25},
          doi = {10.3847/1538-4365/ab6993},
archivePrefix = {arXiv},
       eprint = {1910.04121},
 primaryClass = {astro-ph.CO},
       adsurl = {https://ui.adsabs.harvard.edu/abs/2020ApJS..247...25B}
}

@ARTICLE{2026ApJ...997..320S,
       author = {{Stueber}, Haley R. and {Mantz}, Adam B. and {Allen}, Steven W. and {Flores}, Anthony M. and {Morris}, R. Glenn and {Pan}, Abigail Y. and {Somboonpanyakul}, Taweewat and {Bleem}, Lindsey E. and {Calzadilla}, Michael and {Floyd}, Benjamin and {Hlavacek-Larrondo}, Julie and {McDonald}, Michael and {Sarkar}, Arnab},
        title = "{Deep Chandra Observations of the z = 1.16 Relaxed, Cool-core Galaxy Cluster SPT-CL J2215-3537}",
      journal = {\apj},
         year = 2026,
        month = feb,
       volume = {997},
       number = {2},
          eid = {320},
        pages = {320},
          doi = {10.3847/1538-4357/ae3005},
archivePrefix = {arXiv},
       eprint = {2601.14425},
 primaryClass = {astro-ph.CO},
       adsurl = {https://ui.adsabs.harvard.edu/abs/2026ApJ...997..320S}
}
\bibliographystyle{aasjournal}

\end{document}